\documentclass[twocolumn,twocolappendix]{aastex631}

\usepackage{amsmath}
\usepackage{comment}
\usepackage{breqn}

\submitjournal{ApJ}

\shorttitle{Evolution of the SFRF with Rest-optical Emissions}
\shortauthors{Asada \& Ohta}
\graphicspath{{./}{figures/}}

\begin{document}

\title{Early Growth of the Star Formation Rate Function in the Epoch of Reionization:\\
an Approach with Rest-frame Optical Emissions}

\correspondingauthor{Yoshihisa Asada}
\email{asada@kusastro.kyoto-u.ac.jp}

\author[0000-0003-3983-5438]{Yoshihisa Asada}
\affiliation{Department of Astronomy, Kyoto University \\
Sakyo-ku, Kyoto 606-8502}

\author[0000-0003-3844-1517]{Kouji Ohta}
\affiliation{Department of Astronomy, Kyoto University \\
Sakyo-ku, Kyoto 606-8502}



\begin{abstract}

We present a star formation rate function (SFRF) at $z\sim6$ based on star formation rates (SFRs) derived by spectral energy distribution (SED) fitting on data from rest-frame UV to optical wavelength of galaxies in the CANDELS GOODS-South and North fields.
The resulting SFRF shows an excess compared to the previous estimations by using rest-frame UV luminosity functions (LFs) corrected for the dust attenuation, and is comparable to that estimated from a far-infrared LF.
This suggests that the number density of dust-obscured intensively star-forming galaxies at $z\sim6$ has been underestimated in the previous approach based only on rest-frame UV observations.
We parameterize the SFRF with using the Schechter function and obtain the best-fit parameter of the characteristic SFR (${\rm SFR}^*$) when the faint-end slope and characteristic number density are fixed.
The best-fit ${\rm SFR}^*$ at $z\sim6$ is comparable to that at $z\sim2$, when the cosmic star formation activity reaches its peak.
Together with SFRF estimations with similar approach using rest-frame UV to optical data, the ${\rm SFR}^*$ is roughly constant from $z\sim2$ to $z\sim6$ and may decrease above $z\sim6$.
Since the ${\rm SFR}^*$ is sensitive to the high-SFR end of the SFRF, this evolution of ${\rm SFR}^*$ suggests that the high-SFR end of the SFRF grows rapidly during the epoch of reionization and reaches a similar level observed at $z\sim2$.

\end{abstract}



\section{Introduction} \label{sec:intro}
Galaxies evolve through converting their gas into stars, or star formation activities.
The investigation on the star formation history across the cosmic time is thus essential to understand the galaxy formation and evolution.
To quantitatively evaluate the cosmological evolution of star formation activities in galaxies, it is crucial to examine the star formation rate function (SFRF) of galaxies at a wide range of redshift.

The SFRF is defined as the number density of galaxies written as a function of the star formation rate (SFR) at a given redshift.
Therefore, the SFRF describes the on-going evolution of galaxies at a period of the universe, and the redshift evolution of the SFRF directly gives a picture of cosmological evolution history of galaxies.
Moreover, integrating the SFRF over the SFR provide the star formation rate density (SFRD) at the redshift, which is also useful and important to quantify the stellar mass assembly of galaxies across the cosmic time.

In the high-$z$ ($z\gtrsim4$) universe, the SFRF and/or SFRD is mainly investigated through the rest-frame UV luminosity functions (LFs) \citep[see e.g., a review by][]{madau_cosmic_2014}.
Since the rest-frame UV luminosity is one of the tracers of the SFR of galaxies, the rest-frame UV LF can be converted into the SFRF.
However, the rest-frame UV light can be easily attenuated by dust, since it is necessary to correct for the loss of light due to dust attenuation.
In this correction, the rest-frame UV spectral index $\beta$ ($f_\lambda\propto\lambda^\beta$) is commonly used to estimate the amount of dust attenuation with the IRX-$\beta$ relation (or more directly with $A_{\rm UV}$-$\beta$ relation), which links the spectral slope $\beta$ and the infrared excess defined as ${\rm IRX}=L_{\rm TIR}/L_{\rm UV}$ \citep[see e.g.,][]{meurer_dust_1999}.

On the other hand, recent far-infrared (FIR) observations have been available even at $z\gtrsim4$ \citep[e.g.,][]{rowan-robinson_star_2016,koprowski_evolving_2017}.
FIR observations can probe the dust-obscured star formation activities, and investigations of the SFRF and/or SFRD at $z\gtrsim4$ with FIR observations have been carried out.
However, part of previous studies on FIR observations at $z\gtrsim4$ suggest that the contribution of dust-obscured star formation to the total is negligible and the estimation with rest-frame UV-based approach is well corrected for the dust attenuation \citep[e.g.,][]{koprowski_evolving_2017}, while others suggest the dust-obscured star formation dominates over the dust-unobscured star formation and the rest-frame UV-based approach underestimates the total star formation activities \citep[e.g.,][]{rowan-robinson_star_2016}.
Thus, on the contribution of the dust-obscured star formation at $z\gtrsim4$, no consensus has been reached yet, and an independent examination is desired.

As an independent examination, very recently, investigations on the SFRF and/or SFRD using not only rest-frame UV but also optical data have been reported \citep{asada_star_2021,asada_search_2022,rinaldi_galaxy_2022}.
The rest-frame optical light suffers much less from dust attenuation than rest-frame UV light, thus utilizing rest-frame optical data is expected to derive the properties of galaxies more properly.
Additionally, in the rest-frame optical wavelength, a tracer of the SFR, the H$\alpha$ emission line, is available.
At $z\gtrsim4$, rest-frame optical light including the H$\alpha$ line is redshifted to mid-infreared (MIR) wavelength, and observations with InfraRed Array Camera (IRAC) on Spitzer are the most suitable among the current facilities.
In particular, the H$\alpha$ emission lines from high-$z$ galaxies are strong enough to boost the IRAC broadband photometry \citep[e.g.,][]{yabe_stellar_2009,stark_keck_2013}, and can be recognized through the IRAC color excess.
Therefore, using not only rest-frame UV but also optical data including the H$\alpha$ emission line with Spitzer/IRAC will probe the star formation activities at $z\gtrsim4$ in an independent way of UV-based studies.

As discussed by \citet[A22 hereafter]{asada_search_2022}, to measure the flux of the H$\alpha$ emission line from the IRAC broadband photometry, it is necessary to estimate the contribution from the continuum accurately, and photometry at a band free from any other emission lines is essential.
Particularly, the observation of the continuum at a longer wavelength band than where the H$\alpha$ emission line falls is crucial.
Without an observation at a longer wavelength, the boosted flux and the IRAC color excess due to the emission line can also be interpreted as a red color of stellar continuum or dust reddening.

There are four broadband filters on IRAC, 3.6, 4.5, 5.8, and 8.0 $\mu$m band.
To observe both the H$\alpha$ emission line and the continuum emission at longer wavelength with IRAC, the target redshift must be $z\sim4.5$ ($3.9\lesssim z \lesssim4.9$), $5.8$ ($5.1\lesssim z \lesssim6.6$), or $7.8$ ($6.9\lesssim z \lesssim8.6$), where the H$\alpha$ emission line falls into the 3.6, 4.5, and 5.8 $\mu$m band, respectively.
At $z\sim4.5$, an SFRF is obtained using not only rest-frame UV but also optical data by the spectral energy distribution (SED) fitting method taking the H$\alpha$ emission into account by \citet[A21 hereafter]{asada_star_2021}.
The SFRF at $z\sim4.5$ obtained with this approach shows an excess compared to that estimated from the rest-frame UV-based approach, and suggests the (dust-obscured) intensive star formation may play a more important role in the evolution of galaxies at $z\sim4.5$ than previously expected.
At $z\sim7.8$, A22 obtained a constraint on the SFRF and SFRD through the H$\alpha$ LF derived from the nondetection of H$\alpha$ emission line in the 5.8 $\mu$m band.
However, the SFRF at $z\sim5.8$ has not been investigated with this approach yet, and the redshift evolution of the SFRF obtained with the approach at $z\gtrsim4$ is not known.

In this work, we aim at investigating the SFRF at $z\sim5.8$ using rest-frame UV and optical data including the H$\alpha$ emission line.
At this redshift, the H$\alpha$ emission line is observed with IRAC 4.5 $\mu$m band, and the continuum emission is observed with the 5.8 and 8.0 $\mu$m band.
Thus, we utilize the photometric data with all the four broadbands on IRAC, and perform SED fitting to obtain the SFRF.
In SED fitting, as similar to A21, we extensively examine the assumptions on the model SED including various SFHs, dust attenuation laws, and a two-component model that consists of a young star-forming population and old quenched population, and we take the uncertainty of these model assumptions into account in the resulting SFRF.
In addition, we exploit the resulting SFRF at $z\sim5.8$ and other recent estimations of the SFRF at $z\gtrsim4$ with rest-frame optical emissions, and examine the redshift evolution of the SFRF and SFRD at high-$z$ universe with this independent approach.

This paper is structured as follows.
In Section \ref{sec:data}, we describe the data and the sample selection used in this work.
In Section \ref{sec:estimate}, we derive the physical parameters of the sample galaxies through SED fitting, and we obtain the SFRF using the result of SED fitting in Section \ref{sec:sfrf}.
In Section \ref{sec:implication}, we discuss the implications from the resulting SFRF including the redshift evolution of the SFRF and SFRD at $z\gtrsim4$.
In Section \ref{sec:discussion}, we give further discussions on results in this work, and we summarize this paper in Section \ref{sec:summary}.
Throughout this paper, all magnitudes are quoted in the AB system \citep{oke_secondary_1983}, and we assume the flat cosmological parameters of $H_0=70\ {\rm km\ s^{-1}\ Mpc^{-1}}$, $\Omega_m=0.3$. and $\Omega_\Lambda=0.7$.

\section{Data and Sample Selection} \label{sec:data}
Among surveys carried out with Spitzer, the Great Observatories Origins Deep Survey (GOODS) fields are one of the deepest and widest fields.
Additionally, part of these observations were conducted during the cryogenic mission and thus observations with IRAC 5.8$\mu$m and 8.0$\mu$m band are available, which are essential to achieve our goals.
We focus on both the GOODS-South and -North fields to maximize the survey volume, and use the photometric catalog in the Cosmic Assembly Near-infrared Deep Extragalactic Legacy Survey (CANDELS) program.

\subsection{GOODS-South field}
In GOODS-South field, we use a photometric catalog given by \citet{guo_candels_2013}, which covers a wavelength range from UV to MIR.
Here we briefly summarize the method for the source extraction and photometry in the catalog, but we recommend readers to refer \citet{guo_candels_2013} for details.
The catalog includes multiwavelength band photometry consisting of observations with the Advanced Camera for Surveys (ACS) and Wide Field Camera 3 (WFC3) on Hubble Space Telescope (HST), IRAC on Spitzer, and several ground-based observatories.
The optical data are composed of observations with F435W, F606W, F775W, F814W and F850LP bands on HST/ACS.
The near-infrared (NIR) data are composed of observations with HST/WFC3 F098M, F105W, F125W and F160W bands.
The MIR data contain observations with Spitzer/IRAC 3.6$\mu$m, 4.5$\mu$m, 5.8$\mu$m, and 8.0$\mu$m bands.
From ground-based facilities, the catalog also contains VLT/VIMOS and CTIO/MOSAIC $U$-band data and VLT/ISAAC and VLT/HAWK-I $K_s$-band data.

Photometry on HST images was performed using {\sc SExtractor}'s dual-image mode.
The combined max-depth mosaic in the F160W band was used for source detection, and photometry was performed for point-spread-function (PSF)-matched images.
On ground-based and Spitzer images, photometry was made through {\sc TFIT} \citep{laidler_tfit_2007}.
In this work, we use all the photometry in the 17 bands except for the CTIO $U$-band data from this photometry catalog since this band was revealed to have a red leak \citep{guo_candels_2013}.

We utilize the CANDELS photometric redshift catalog \citep{dahlen_critical_2013}.
For the entire sources in the photometric catalog, photometric redshift is derived using a hierarchical Bayesian approach that combines the probability distribution functions (PDFs) estimated from several manners.

\subsection{GOODS-North field}
In GOODS-North field, we use a catalog given by \citet{barro_candelsshards_2019}, which contains photometry from UV to far-infrared.
The catalog contains not only the photometry but also photometric redshifts and other stellar parameters, but, in this work, we only use the photometry from UV to MIR and the photometric redshift.
The catalog includes mutiwavelength photometry on observations with ACS and WFC3 on HST, IRAC on Spitzer, and several ground-based observatories.
For optical data, observations with HST/ACS F435W, F606W, F775W, F814W, and F850LP bands are available.
The NIR data consist of observations with HST/WFC3 F105W, F125W, F140W, and F160W bands.
The MIR data are composed of observations with Spitzer/IRAC 3.6$\mu$m, 4.5$\mu$m, 5.8$\mu$m, and 8.0$\mu$m bands.
From ground-based facilities, KPNO/Mosaic and LBT/LBC $U$-band, Subaru/MOIRCS $K_s$-band, and CFHT/Megacam $K$-band observations are included.
This catalog also includes optical medium-band data that covers $\lambda=2500$ to $9500$ \AA\ from the GTC SHARDS survey and WFC3 IR spectroscopic observations with the G102 and G141 grisms, but none of our sample galaxies are detected in these observations (c.f., sample selection in Section \ref{subsec:sample}) and we do not use data from these observations in the following analysis.

Photometry in this catalog was carried out in a similar way as in the previous CANDELS catalogs including that by \citet{guo_candels_2013} in GOODS-South field, which is introduced in the previous subsection.
The source detection was performed in the F160W-band image, and photometry on all the HST band images was made for PSF-matched images with {\sc SExtractor}'s dual-image mode.
For the ground-based and Spitzer/IRAC observations, photometry was conducted through {\sc TFIT}.

The catalog also contains photometric redshifts for the entire sources in the catalog.
The photometric redshifts are also derived in a similar approach as in the previous CANDELS catalogs: the photometric redshift is derived by combining the PDFs estimated from several manners.

\subsection{Sample Selection} \label{subsec:sample}
To make a sample of galaxies at $5.09<z<6.62$ whose H$\alpha$ emissions are redshifted into the IRAC 4.5$\mu$m band, we set the sample selection criteria as follows.

First, we select galaxies based on the photometric redshift.
Specifically, we extract galaxies whose best-estimated photometric redshift $z_{\rm best}$ is in our target range and whose relative uncertainty on the photo-$z$ estimation is less than 5\%:
\begin{equation}
    5.09 < z_{\rm best} < 6.62, 
\end{equation}
and
\begin{equation}
    \frac{\Delta z}{1+z} < 0.05.
\end{equation}
Next, we exclude X-ray sources to remove active galactic nuclei (AGNs) from our sample.
In GOODS-S field, we eliminate AGNs identified by \citet{hsu_candelsgoods-s_2014}.
The X-ray sensitivity limits of this AGN catalog are typically $3.2\times10^{-17}$, $9.1\times10^{-18}$, and $5.5\times10^{-17}$ erg s$^{-1}$ cm$^{-2}$ for the full (0.5-8 keV), soft (0.5-2keV), and hard (2-8keV) bands, respectively.
These correspond to luminosity limits of $1.2\times10^{43}$, $3.4\times10^{42}$, and $2.0\times10^{43}$ erg s$^{-1}$ with assuming the redshift $z=5.8$, respectively.
In GOODS-N field, we eliminate X-ray sources identified by \citet{alexander_chandra_2003}.
The sensitivity limits are about $2.5\times10^{-17}$ and $1.4\times10^{-16}$ erg s$^{-1}$ cm$^{-2}$ for soft (0.5-2.0 keV) and hard (2-8 keV) bands, which correspond to luminosity limits of $9.2\times10^{42}$ and $5.2\times10^{43}$ erg s$^{-1}$ by the assumption of redshift, respectively.

Among the 34,930 (35,445) objects in the CANDELS GOODS-S (GOODS-N) catalog, 300 (254) galaxies pass the criteria above (we refer them as "phot-$z$ sample").
We apply additional cuts to secure the reliability on SEDs of galaxies in our sample and to make a robust estimation of their physical properties by SED fitting.
As introduced in Section \ref{sec:intro}, the detection of the rest-frame optical continuum is important to estimate the contribution of the continuum in 4.5$\mu$m band.
Thus, we require the signal-to-noise ratio (S/N) both in IRAC 5.8$\mu$m and 8.0$\mu$m bands larger than 3.
Among the 554 ($=300+254$) galaxies in the phot-$z$ sample, 23 galaxies have ${\rm S/N}>3$ both in IRAC 5.8$\mu$m and 8.0$\mu$m bands.
We also remove galaxies that are heavily blended with the neighboring objects in the IRAC images.
Although the photometry in IRAC bands was performed with a deblending tool, {\sc TFIT}, heavy contamination can lead to a systematic uncertainty in the photometry.
To this end, we eliminate all the galaxies whose separation from the nearest object in the photometric catalog is $<2^{\prime\prime}$, considering the FWHMs of IRAC images are $\sim2^{\prime\prime}$.
Finally, we conduct visual inspections on the galaxies that meet criteria above, and obtain a sample of galaxies that contains nine galaxies (six from GOODS-S and three from GOODS-N).
A list of these nine galaxies is shown in Table \ref{tab:obj_list}, and the distribution of apparent magnitudes at $H_{160}$ and 5.8$\mu$m band is shown in the lower left panel of Figure \ref{fig:mags}.
In Figure \ref{fig:cutouts}, postage stamps of galaxies in the final sample are shown.
In Appendix \ref{apd:physical_prop}, the SEDs of the nine galaxies are also shown.

\begin{figure}[tpb]
\centering
\includegraphics[width=0.9\columnwidth, angle=0]{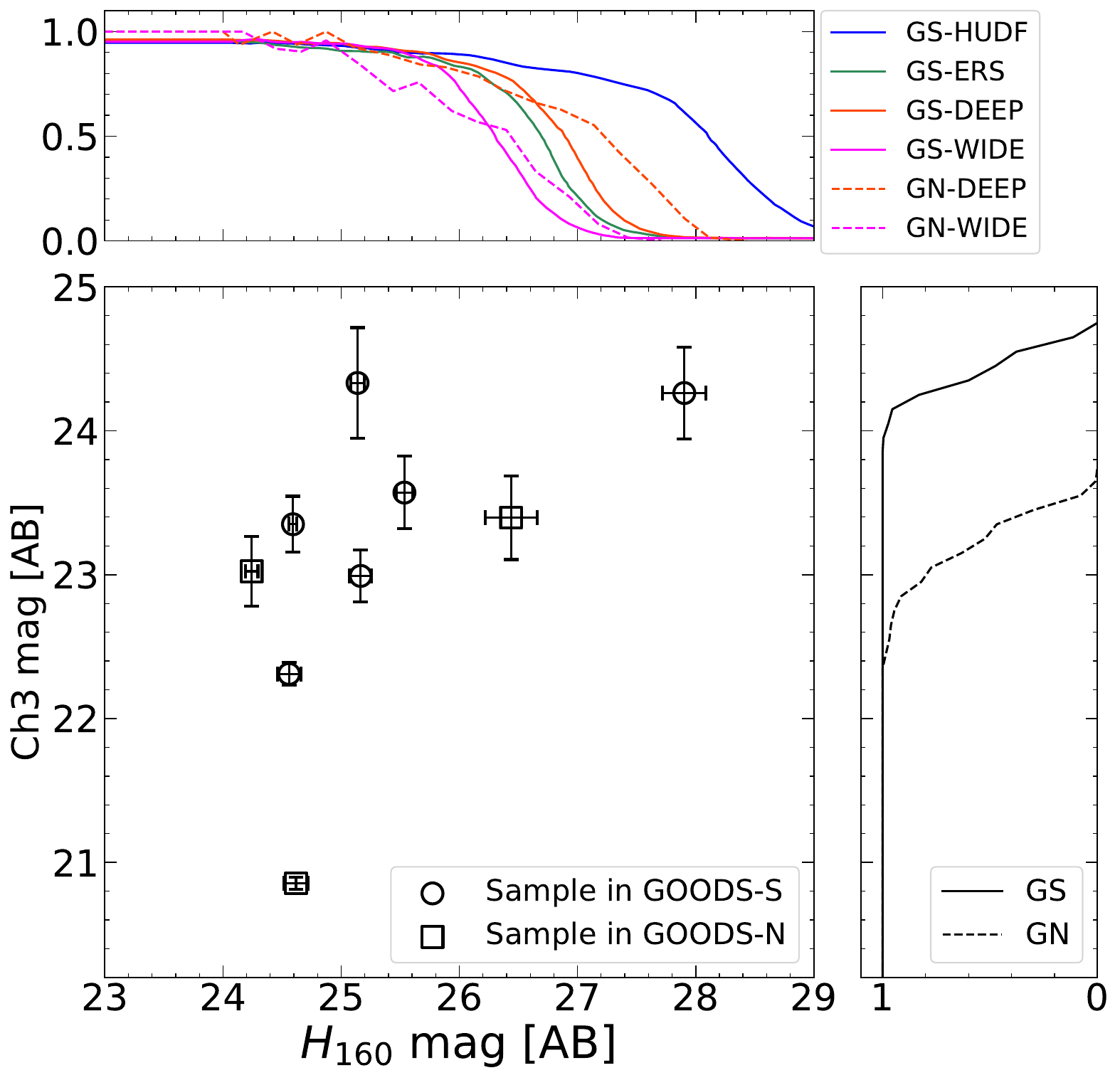}
\caption{Top: detection rates at the $H_{160}$ band in GOODS-S and -N fields.
The different linestyle shows different fields, and different color shows different regions in a field.
Lower left: the distribution of apparent magnitudes at $H_{160}$ versus 5.8$\mu$m band of galaxies in the final sample.
The circles (squares) represent galaxies in the GOODS-S (-N) field.
Lower right: normalized cumulative histogram of the $3\sigma$ limiting magnitudes in the 5.8$\mu$m (Ch3) band, which effectively shows the detection rate in the band. (see text for details).
}
\label{fig:mags}
\end{figure}

\begin{figure*}[tpb]
\centering
\includegraphics[width=2.0\columnwidth, angle=0]{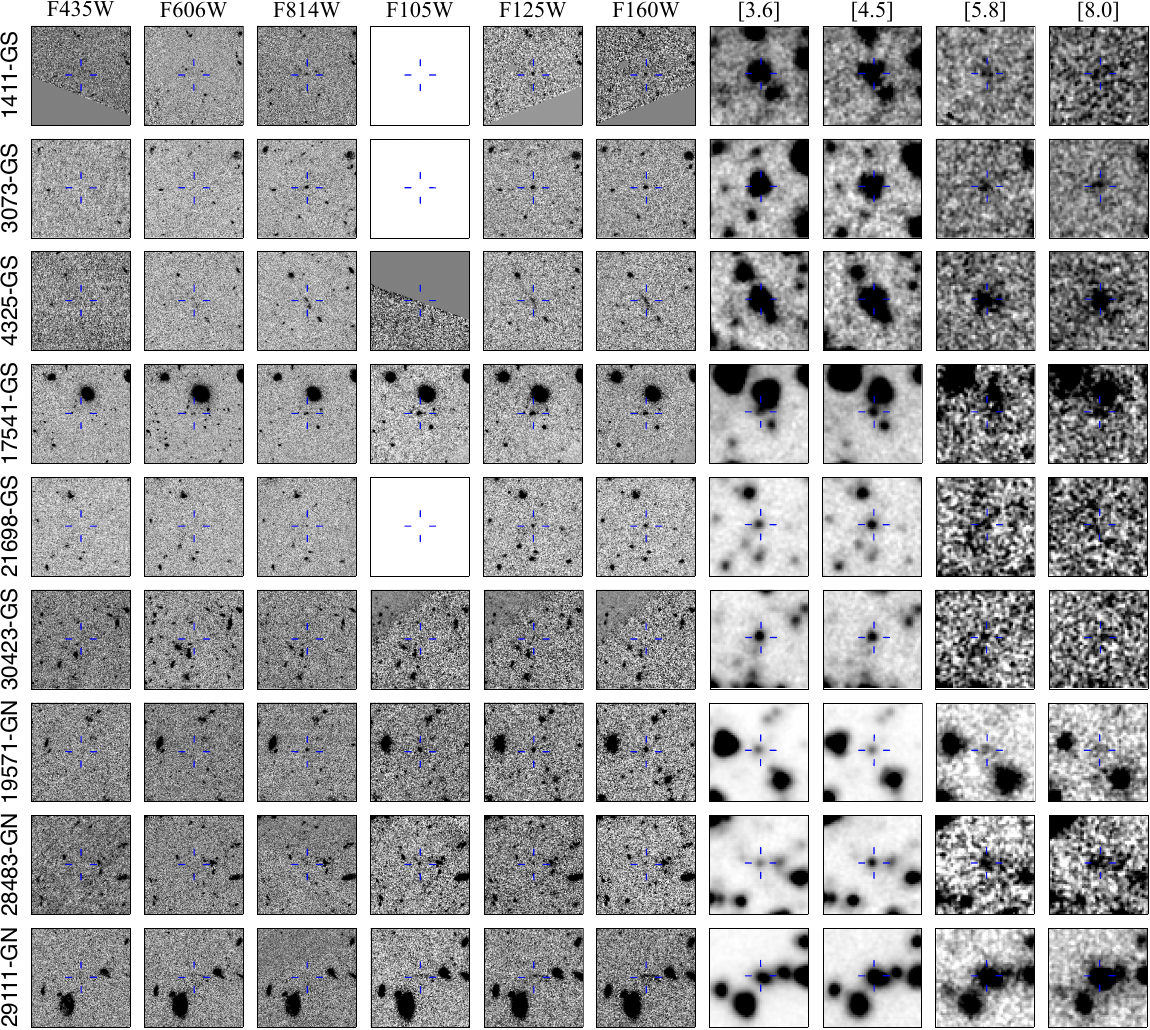}
\caption{Postage stamps ($16^{\prime\prime}\times16^{\prime\prime}$) of galaxies in the final sample.
From left to right, images in ACS F435W, F606W, F814W, WFC3 F105W, 125W, F160W, IRAC 3.6 $\mu$m, 4.5 $\mu$m, 5.8 $\mu$m, and 8.0 $\mu$m bands are shown for each galaxy (top to bottom).
The HST and Spitzer images are taken from publicly available science images created by the Hubble Legacy Fields project and by the GOODS Reionization Era wide-Area Treasury from Spitzer project \citep{stefanon_spitzerirac_2021}, respectively.
}
\label{fig:cutouts}
\end{figure*}

\begin{deluxetable*}{lccccccc}
\tablenum{1}\label{tab:obj_list}
\tablecaption{Galaxies in the final sample}
\tablewidth{0pt}
\tablehead{
\colhead{Name\tablenotemark{a}} & \colhead{R.A.} & \colhead{Dec.} & \colhead{$H_{160}$} &
\colhead{[5.8]} & \colhead{[8.0]} & \colhead{$\beta$\tablenotemark{b}} & \colhead{Phot-$z$} \\
\colhead{} & \colhead{(J2000)} & \colhead{(J2000)} & \colhead{(mag)} &
\colhead{(mag)} & \colhead{(mag)} & \colhead{} & \colhead{} 
}
\decimalcolnumbers
\startdata
1411-GS & 53.2477072 & -27.9083695 & 25.16$\pm$0.09 & 22.99$\pm$0.18 & 23.47$\pm$0.26 & -1.07$\pm$0.15 &5.46$\pm$0.14 \\
3073-GS & 53.0788207 & -27.8840976 & 24.59$\pm$0.03 & 23.35$\pm$0.19 & 22.94$\pm$0.14 & -1.78$\pm$0.06 &5.56$\pm$0.05\tablenotemark{c}\\
4325-GS & 53.1294110 & -27.8714760 & 24.56$\pm$0.10 & 22.31$\pm$0.08 & 21.85$\pm$0.05 & -0.66$\pm$0.17 &5.14$\pm$0.16\\
17541-GS & 53.1708041 & -27.7622288 & 25.14$\pm$0.06 & 24.33$\pm$0.38\tablenotemark{d} & 24.18$\pm$0.38\tablenotemark{d} & -1.76$\pm$0.07 &5.42$\pm$0.05\\
21698-GS & 53.0235059 & -27.7252509 & 25.54$\pm$0.07 & 23.57$\pm$0.25 & 23.67$\pm$0.29 & -0.30$\pm$0.33 &5.95$\pm$0.21\\
30423-GS & 53.1547494 & -27.8064591 & 27.90$\pm$0.18 & 24.26$\pm$0.32 & 24.45$\pm$0.39\tablenotemark{d} & -0.33$\pm$0.33 &5.28$\pm$0.16\\
19571-GN & 189.5032440 & 62.2738825 & 24.24$\pm$0.05 & 23.02$\pm$0.24 & 23.28$\pm$0.28 & -1.36$\pm$0.08 &5.45$\pm$0.04\\
28483-GN & 189.1723816 & 62.1570918 & 26.44$\pm$0.22 & 23.40$\pm$0.29 & 22.98$\pm$0.21 & -1.69$\pm$0.36 &5.27$\pm$0.19\\
29111-GN & 188.9601633 & 62.1784043 & 24.62$\pm$0.10 & 20.86$\pm$0.04 & 20.38$\pm$0.03 & -0.35$\pm$0.33 &5.67$\pm$0.13\\
\enddata
\tablenotetext{a}{
Name represents the ID given in the CANDELS catalogs and the field where the galaxy is located.}
\tablenotetext{b}{
The rest-frame UV spectral slope $\beta$ is directly measured from the rest-frame UV photometry.}
\tablenotetext{c}{
The spectroscopic redshift is available only for 3073-GS, $z_{\rm spec}=5.563$.}
\tablenotetext{d}{
Some sources have a relatively large magnitude errors on 5.8- and 8.0-$\mu$m band photometry although they pass the S/N cut in the sample selection.
This is because we used the limiting magnitudes as defined by averaged rms values given in the catalog for the S/N calculation while the magnitude errors in this table are nominally taken from the photometric errors in the catalog.}
\end{deluxetable*}

\section{Estimation of the Physical Properties}\label{sec:estimate}
\subsection{SED fitting}\label{subsec:SEDfit}
To estimate the physical properties including the SFR, we first carry out SED fitting on the galaxies in our sample.
We use the population synthesis code {\sc P\'{e}gase.}3 \citep{fioc_pegase3_2019} to make model SEDs.
This code includes the nebular emissions in the model spectrum, and follows the chemical evolution in a self-consistent way, which is used to calculate the metallicity of the interstellar medium and the next generation stars\footnote{We do not consider the presence of inflowing gas into the system in the SED modeling, which corresponds to the so-called {\it closed-box} model.}.
We adopt Chabrier03 IMF \citep{chabrier_galactic_2003} with a mass range of $0.08$-$120\ M_\odot$, and the attenuation by intergalactic neutral hydrogen is modeled following the prescription given by \citet{madau_radiative_1995}.
The internal dust attenuation is modeled with Calzetti law \citep{calzetti_dust_2000} and Small Magellanic Cloud \citep[SMC;][]{pei_interstellar_1992} law.
We assume $R_V=4.05$ and $R_V=2.93$ for Calzetti and SMC law, respectively.
As for the star formation history (SFH), we consider seven models; constant star formation (CSF), exponentially declining ($\propto e^{-t/\tau}$), and delayed exponential ($\propto te^{-t/\tau}$) models with the $e$-folding time of $\tau=10,\ 100, 1000$ Myr.

Under these assumptions, we search for the best-fitting model SED with the free parameters of age and color excess $E(B-V)$ by $\chi^2$ minimization.
We allow the age of galaxies to vary with $\sim60$ steps from 1 Myr to the age of the universe at the redshift where each galaxy is located.
The color excess is taken from 0.0 to 0.8 mag at an interval of 0.01 mag. The ratio between stellar and nebular attenuation is assumed to be 1, since this ratio is suggested to be about unity in the high-$z$ universe \citep[e.g.,][]{cullen_mass-metallicity-star_2014,reddy_mosdef_2015,theios_dust_2019}.
Note that, we find that the best-fitting $E(B-V)$ is larger than 0.8 for only one galaxy (29111-GN), and thus we allow $E(B-V)$ to vary up to 1.5 for this galaxy.
We do not use the photometry at the wavelength shortward to Ly$\alpha$ in calculating the $\chi^2$.
Consequently, we obtain 14 best-fitting model SEDs for each galaxy in the final sample; two types of assumptions on dust attenuation law and seven on SFHs, and each of them gives a respective set of best-fitting physical parameters including age, $E(B-V)$, stellar mass $M_\star$, and SFR.

\subsection{Two-component model SED fitting}\label{subsec:two-comp}
As discussed by A21, the SED fitting on high-$z$ galaxies with a red SED from rest-frame UV to optical wavelength range can result in a significantly dusty and highly star-forming solution, but the red color in SEDs can be explained not only by the dust attenuation but also by the aging of stellar population.
A21 showed that considering a two-component model that consists of old quenched stellar population and young star-forming population in SED fitting can reduce the best-fitting $E(B-V)$ and ${\rm SFR}$ particularly for highly star-forming solutions, leading to an effect on the resulting SFRF.
We thus also consider such a two-component model.
Here, we adopt a similar approach as that by A21.
For simplicity, we assume there is no dust attenuation in the old stellar population, and fix the spectrum of the old stellar component to that is expected for the reddest in the wavelength range of rest-frame UV to optical within the age of the universe.
Since the age of the universe at $z\sim5.8$ is $\lesssim1$ Gyr, the age of the old component is fixed to 500 Myr\footnote{The onset of the star formation of 500 Myr old population at $z\sim5.8$ is $z\sim10$.}.
We compare the continuum spectra at the age of 500 Myr with four SFHs; instantaneous burst and CSF with a duration of $\Delta \tau =10, 50, 100$ Myr, and find that the CSF with $\Delta \tau=100$ Myr gives the reddest spectrum, which is consistent with the result shown by A21\footnote{Note that the color of continuum spectra can be affected by the chemical evolution (see appendix in A21).}.
Thus, we adopt the spectrum from the CSF model with $\Delta\tau=100$ Myr aged 500 Myr as the old stellar population spectrum.

Using this spectrum, we subtract the old population SED with a fraction of $f_{\rm old}$ described below from the observed SED, and perform SED fitting as described in Section \ref{subsec:SEDfit} to the residual, which gives the best-fitting SED of "young star-forming population".
Here, we allow the contribution of the old stellar population to the whole SED to vary; specifically, we allow the fraction of flux in the 5.8$\mu$m band by the old stellar population to the total (observed) flux ($f_{\rm old}$) to be changed from 0 to 0.95.
Thus, in this two-component model, free parameters and the assumptions for the young star-forming population are the same as those in the normal, one-component model that is described in Section \ref{subsec:SEDfit}, and one additional free parameter $f_{\rm old}$ is considered.
We obtain 14 best fit SEDs and the number of sets of best-fitting physical parameters for each galaxy with this two-component SED fitting as well.
The $M_\star$ obtained in this two-component model is the estimation for the whole system, and the SFR, age, and $E(B-V)$ are those for the young component.
Note that, since the SFR of the old component is assumed to be zero, the SFR obtained here can be regarded as the estimation for the whole system.

Among the 14 sets of best-fitting physical parameters from this two-component model, in the following analysis, we only use results that meet following criteria:
\renewcommand{\theenumi}{\Roman{enumi}}
\begin{enumerate}
    \item\label{Criteria1} The best-fitting age for the young population is less than 400 Myr.
    \item\label{Criteria2} $f_{\rm old}$ is larger than 0.5.
\end{enumerate}
In this model, the old population is assumed to be quenched 400 Myrs ago, so the young component must be younger than 400 Myr (\ref{Criteria1}).
If the best-fit $f_{\rm old}$ is small, that means old stellar population is almost negligible and/or observed SED is well fitted without such an old population (\ref{Criteria2}).

\subsection{Physical parameter estimation}\label{subsec:physpara}
From the results in Section \ref{subsec:SEDfit} and \ref{subsec:two-comp}, we obtain 14 or less sets of estimations\footnote{For each attenuation law, seven estimations from the one-component SED fitting in Section \ref{subsec:SEDfit} and seven or less estimations from the two-component fitting in Section \ref{subsec:two-comp}.} of physical parameters (i.e., age, $E(B-V)$, SFR, and $M_\star$) for each attenuation law for each galaxy in the final sample.

Using these sets of estimations, we aim at deriving the best estimation of each physical parameter taking into account the goodness-of-fit of the model SEDs under the assumption of the attenuation law for each galaxy.

First, we evaluate the goodness-of-fit for each of the 14 or less estimations using the Bayesian Information Criterion (BIC) that is defined as
\begin{equation}
    {\rm BIC} = \chi_{\rm min}^2 + q\ln(m).
\end{equation}
where $\chi_{\rm min}^2$ is the minimized $\chi^2$ given by the best fit model SED, $q$ is the number of free parameters ($q=2$ for results from Section \ref{subsec:SEDfit} and $q=3$ for results from Section \ref{subsec:two-comp}), and $m$ is the number of independent observations.
The smaller value of BIC means the model SED fits better to the observed SED, and the likelihood $\mathcal{L}$ of the model SED given the observed data is $\mathcal{L}\propto\exp(-{\rm BIC}/2)$.
Naively, the estimation that gives the smallest BIC is the most likely, and thus one may think adopting the physical parameter estimation with the smallest BIC (which gives the maximum likelihood) is reasonable.
However, the maximum likelihood value is not distinctively larger than those in other cases, and the difference in physical parameter estimations is not negligible between the maximum likelihood model and other models with comparable likelihood (c.f., Figure \ref{fig:bootstrap_1411} in Appendix \ref{apd:Example_bootstrap}).
Thus adopting only the estimation with the maximum likelihood and discarding all the other estimations seem to be risky.
To take this into account, for each physical parameter, we calculate the weighted mean and weighted standard deviation using $\exp(-{\rm BIC}/2)$ as the weight, and use them as the best estimations and its uncertainties for each physical parameters.
We derive weighted means and weighted standard deviations for each of attenuation law assumptions (Calzetti law or SMC law).
This gives the physical parameter estimations under the assumption of attenuation law.
We also derive these values using all the set of best-fit SEDs, including both of Calzetti and SMC assumption, to obtain a best estimation taking the uncertainty of dust attenuation law into account (we call this estimation as "Calzetti+SMC estimation").

As a result, for each physical parameter, we derive three estimations and the associated uncertainties: Calzetti, SMC, and Calzetti+SMC estimations.
In Appendix \ref{apd:Example_bootstrap}, we show an example of this physical parameter estimation described in this subsection for one galaxy in the final sample (1411-GS).

\begin{figure*}[tpb]
\centering
\includegraphics[width=2.0\columnwidth, angle=0]{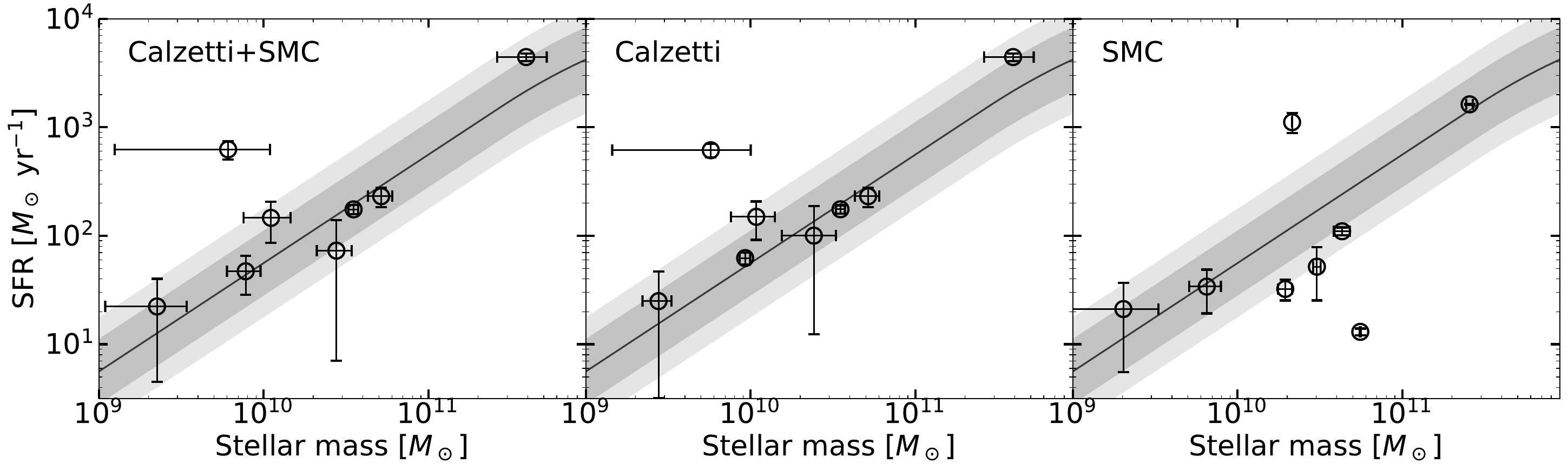}
\caption{The stellar mass versus SFR diagram for galaxies in the final sample.
The best estimations of the stellar mass and SFR obtained in Section \ref{subsec:physpara} using Calzetti+SMC, Calzetti, SMC estimation are used to plot in the left, middle, and right panel, respectively.
The black solid line denotes the MS relation at $z\sim5.8$ predicted from the redshift evolution of the MS obtained by \citet{schreiber_herschel_2015}.
The thick (thin) shaded region show a typical scatter of $\pm0.3$ ($\pm0.5$) dex from the MS.
}
\label{fig:MstarvsSFR}
\end{figure*}

In Figure \ref{fig:MstarvsSFR}, we show the resulting stellar mass against SFR obtained with Calzetti+SMC, Calzetti, and SMC estimation in the left, middle, and right panel, respectively.
For comparison, we show a main sequence (MS) relation at $z\sim5.8$ predicted by the redshift evolution of the MS relation by \citet{schreiber_herschel_2015}.
\citet{schreiber_herschel_2015} investigated the MS relation from $z=0$ to $z=4$ and derived the redshift evolution.
Here we extrapolate the evolution up to $z\sim5.8$ to plot the MS relation in the figure.
We can see that most galaxies in the final sample are aligned with the MS relation without a few exceptions; one starburst and one or two quiescent galaxies.
Thus, the final sample mainly contains (massive) MS galaxies at this redshift and a few starburst or quiescent galaxies.
Note that the SED of 30423-GS is well fitted with quiescent solutions, and its SFR value is so low that the galaxy is missing in Figure \ref{fig:MstarvsSFR} and \ref{fig:Ch3vsSFR}.
The MS relation at $z\sim5.8$ does not differ significantly between literature, and the consequence here persists even if we compare with MS relations obtained in other studies \citep[e.g.,][]{speagle_highly_2014,santini_star_2017,popesso_main_2022}.
In Appendix \ref{apd:physical_prop}, we show the results of SED fittings and the resulting estimations of physical parameters for all the galaxies in the final sample.

The methodology described in this subsection is essentially for estimating physical parameters taking various SFHs into account.
This method is simple and easy to be applied even with results from SED fitting code assuming parameterized SFH, but the statistical validity may be unclear.
Thus, it is worth comparing with other non-parametric SFH SED fitting approaches.
We use a SED fitting code {\tt dense basis} \citep{iyer_2017,iyer_2019}, which assumes non-parametric SFH, and examine how it affects the results in this paper.
Results in this paper persist when we use {\tt dense basis} estimations instead of the weighted-mean estimations (see Appendix \ref{apd:DB}).

\section{Star Formation Rate Function}\label{sec:sfrf}
In this section, we describe the method for deriving the SFRF using the result in Section \ref{sec:estimate}.
We use a similar approach to derive the SFRF as by A21, but slightly modified to apply to the data set in this work.

\subsection{Method for deriving SFRF}\label{subsec:method_sfrf}
It is necessary to correct for the incompletness of the sample to properly derive the SFRF.
As similar to A21, we consider three factors causing the incompleteness: (i) detection in the $H_{160}$ band, (ii) S/N cut in the 5.8 $\mu$m and 8.0$\mu$m band, and (iii) elimination of the blended objects in the IRAC images.

To correct for (i), the detection rate in the $H_{160}$-band image for each field is necessary.
In GOODS-S, \citet{duncan_mass_2014} derived the detection rate of the same CANDELS GOODS-S catalog as we use in this work by performing mock observations, and we use the result.
The solid lines in the top panel of Figure \ref{fig:mags} show the detection rate as a function of $H_{160}$ magnitude.
\citet{duncan_mass_2014} obtained the detection rate by dividing the GOODS-S field into four subregions according to the exposure time; HUDF, ERS, DEEP, and WIDE.
In GOODS-N, we estimate the $H_{160}$-band detection rate from the differential number counts for objects in the catalog following the prescription by \citet{guo_candels_2013}.
\citet{barro_candelsshards_2019} derived the differential number density and the best-fit power law function dividing the field into two subregions (DEEP and WIDE).
We use the result and estimate the detection rate in the faint region by calculating the ratio of the observed number density to that predicted by the best-fit power law.
The dashed lines in the top panel of Figure \ref{fig:mags} show the resulting detection rate.

As for the incompleteness due to (ii), since the exposure times in IRAC 5.8$\mu$m and 8.0$\mu$m band are also inhomogeneous, we correct for the effect of S/N cut considering this inhomogeneous depth of these images.
To this end, we calculate the 3$\sigma$ limiting magnitudes in the 5.8$\mu$m band at the locations of galaxies in the phot-$z$ sample (see Section \ref{subsec:sample} for the definition of the phot-$z$ sample), and make the normalized cumulative histogram of them in each GOODS-S and -N fields.
In the CANDELS GOODS-S catalog, 1$\sigma$ limiting magnitudes at the locations of the objects are available, and we use them to calculate the 3$\sigma$ limiting magnitudes to make the normalized cumulative histogram.
In the CANDELS GOODS-N catalog, on the other hand, they are not available, thus we use the flux uncertainties of galaxies in the phot-$z$ sample instead.
Since the phot-$z$ sample are composed of galaxies at $z\sim5.8$ without AGN activities, their photometric errors are expected to be background limited\footnote{The typical $H_{160}$ magnitude of galaxies in the phot-$z$ sample is $\sim26.13$ mag.}.
As discussed by A21, we can use the normalized cumulative histogram of the $3\sigma$ limiting magnitudes in the 5.8 $\mu$m band of galaxies in the phot-$z$ sample to correct for the incompleteness due to S/N cut in the 5.8$\mu$m band.
In the sample selection, we applied S/N cut not only in 5.8$\mu$m band but also in 8.0$\mu$m band, so one might think an additional correction is required.
However, firstly, the limiting magnitudes at 5.8$\mu$m and 8.0$\mu$m band are almost identical.
Secondly, the observed color of $[5.8]-[8.0]$ is typically 0 or more (c.f. Figure \ref{fig:SED} in Appendix \ref{apd:physical_prop}).
Finally, the exposure maps in 5.8$\mu$m and 8.0$\mu$m band are similar to each other, and regions with a deep observation in 5.8$\mu$m band is also expected to have a deep observation in 8.0$\mu$m band.
Considering these factors, galaxies that are detected in the 5.8$\mu$m band with S/N$>3$ should be detected also in the 8.0$\mu$m band (c.f., Figure \ref{fig:cutouts}), and thus an additional correction should not be necessary.
In the right panel of Figure \ref{fig:mags}, the resulting normalized cumulative histograms in the GOODS-S and -N fields are shown.

We correct for the incomleteness due to (iii) using the fraction ($f^{\rm sel}$) of isolated objects both in 4.5- and 5.8-$\mu$m band IRAC images at $z\sim5.8$ as similar to A21.
To evaluate $f^{\rm sel}$, we first randomly pick out $\sim200$ galaxies from the phot-$z$ sample, and categorize them into the following three groups by visual inspections; galaxies that are detected in the 5.8$\mu$m band image and not blended with the neighbors (group A), galaxies that are detected in the 5.8$\mu$m band image but heavily blended with the neighbors (group B), and objects that are not detected in the 5.8$\mu$m band image.
We then calculate $f^{\rm sel}$ by $\#(A)/(\#(A)+\#(B))$, and obtain result in $f^{\rm sel}\sim0.52$.

Taking these incomleteness factors into account, we can estimate the SFRF $\phi(\dot{M}_\star)\ {\rm [Mpc^{-3}\ dex^{-1}]}$ in the bin from $\dot{M}_{\star,1}$ to $\dot{M}_{\star,2}$ with a bin width of $d(\log \dot{M}_\star)$ as follows;
\begin{equation}\label{eqn:SFRF_calc}
    \phi(\dot{M}_\star) d(\log \dot{M}_\star) = \sum_i^{N_{\rm gal}} \frac{1}{V_i^{\rm eff}}\int_{\dot{M}_{\star,1}}^{\dot{M}_{\star,2}} P_i(\dot{M}_\star) d\dot{M}_\star
\end{equation}
where the subscript $i$ represents each galaxy in the final sample, $N_{\rm gal}$ is the number of galaxies $(=9)$, $P_i( \dot{M}_\star )$ is the PDF of the SFR for the galaxy $i$, and the $V_i^{\rm eff}$ is the effective volume of this survey for the galaxy $i$.
We assume Gaussian profiles for $P_i( \dot{M}_\star )$, with the best estimations and their uncertainties obtained in Section \ref{subsec:physpara} being used as the mean and sigma of the profile.
This effective volume $V_i^{\rm eff}$ can be calculated as
\begin{align}\label{eqn:eff_vol}
    V^{\rm eff}_i =& \notag \\
    \sum_k^{N_{\rm region}} &f^{\rm sel} f^{160}_k(m_{160,i}) f^{\rm ch3}_k(m_{{\rm ch3},i}) \Omega_k \int_{r_{z=5.09}}^{r_{z=6.62}} r^2 dr
\end{align}
where subscript $k$ represents the regions in the fields, $N_{\rm region}$ is the number of regions $(=6)$, $f^{160}_k$ is the detection rate in the $H_{160}$ band as a function of the $H_{160}$-band magnitude (i), $f^{\rm ch3}_k$ is the value in the cumulative histogram of 3$\sigma$ limiting magnitudes as a function of the 5.8$\mu$m-band magnitude (ii), $f^{\rm sel}$ is the fraction of isolated objects in IRAC images (iii), $m_{160,i}$ and $m_{{\rm ch3},i}$ are the observed magnitudes for the galaxy $i$ in the $H_{160}$ and 5.8$\mu$m band, respectively, $\Omega_k$ is the solid angle of the region $k$, and $r_{z=a}$ is the comoving distance from $z=0$ to $z=a$.

\begin{figure*}[tpb]
\centering
\includegraphics[width=2.0\columnwidth, angle=0]{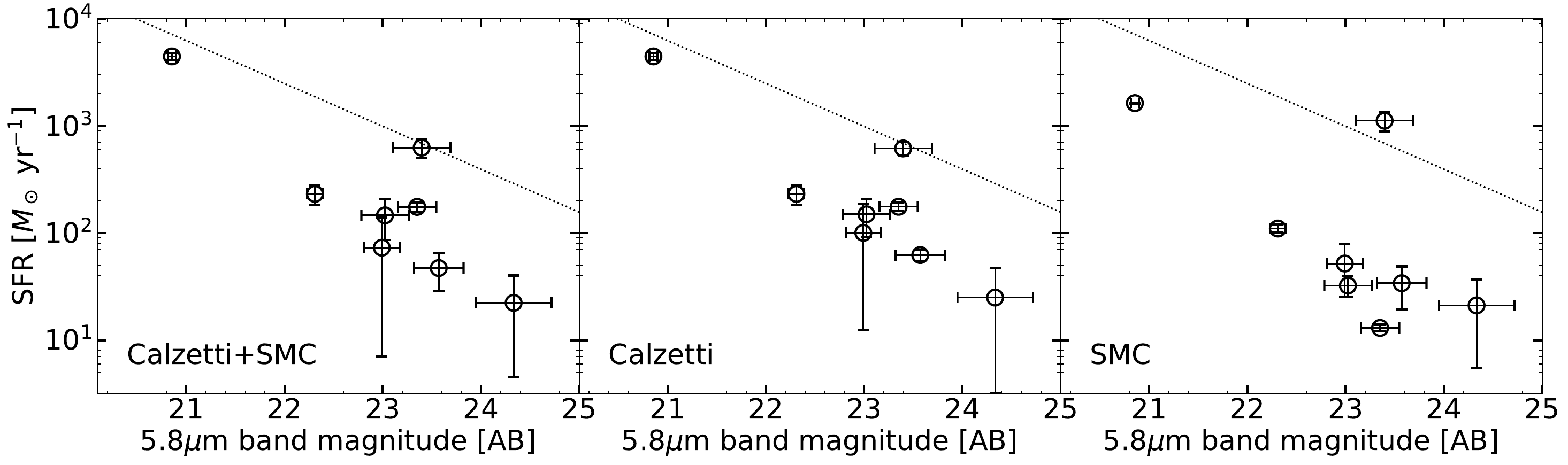}
\caption{The SFRs versus 5.8$\mu$m band magnitudes for galaxies in the final sample.
Left, middle, and right panel shows the diagram with Calzetti+SMC, Calzetti, and SMC estimations, respectively.
The black dotted line represents a guideline that corresponds to the upper limit on the SFR obtained from a maximum value of rest-frame H$\alpha$ EW (${\rm EW(H}\alpha)=4000$\AA), and the region above this line is expected to be forbidden (see the text for details).
}
\label{fig:Ch3vsSFR}
\end{figure*}

We also estimate the completeness limit of the SFR, above which the SFRF obtained with Equation (\ref{eqn:SFRF_calc}) is well-corrected for the incompleteness.
To estimate the completeness limit of the SFR, we show the diagram of 5.8$\mu$m-band magnitude versus the SFR (Figure \ref{fig:Ch3vsSFR}).
We can see a rough anti-correlation between the [5.8] and SFR regardless of the assumption of the dust attenuation.
However, due to the small number of the sample galaxies, it is difficult to estimate the completeness limit only from this distribution.
In the figure, we also plot a guideline that shows the upper limit on the SFR by the black dotted line.
The rest-frame H$\alpha$ equivalent width (EW) cannot be larger than $\sim4000$ \AA\ even if most extreme stellar populations (e.g., PopIII stars) are considered \citep[e.g.,][]{inoue_rest-frame_2011}.
If we assume the 5.8$\mu$m flux density to be the continuum around H$\alpha$ at $z\sim5.8$, we then can obtain an upper limit on the H$\alpha$ luminosity for a given magnitude in the 5.8 $\mu$m band, which leads to an upper limit on the SFR\footnote{Since 5.8$\mu$m corresponds to $\lambda_{\rm rest}\sim8500$ \AA\ in galaxies at $z\sim5.8$, this assumption means we assume a flat continuum spectrum at rest-frame optical ($\lambda_{\rm rest}\sim6500-8500$ \AA).}.
In spite of a rough estimation of the upper limit, this constraint is consistent to the SFRs derived in this work.
Considering this upper limit and the overall distribution of galaxies in the [5.8]-SFR diagram, galaxies with SFR $\gtrsim10^{2.5}\ M_\odot\ {\rm yr}^{-1}$ should be brighter than $\sim24.5$ mag in the 5.8$\mu$m band, but those with SFR $\lesssim10^{2.5}\ M_\odot\ {\rm yr}^{-1}$ can be fainter than $\sim24.5$ mag in the band.
Thus, the limiting magnitude of $\sim24.5$ mag in the 5.8$\mu$m band (c.f. lower right panel in Figure \ref{fig:mags}) corresponds to the completeness limit of SFR $\sim10^{2.5}\ M_\odot\ {\rm yr}^{-1}$.

\subsection{Star Formation Rate Function}\label{subsec:SFRF}
\begin{figure*}[tpb]
\centering
\includegraphics[width=2.0\columnwidth, angle=0]{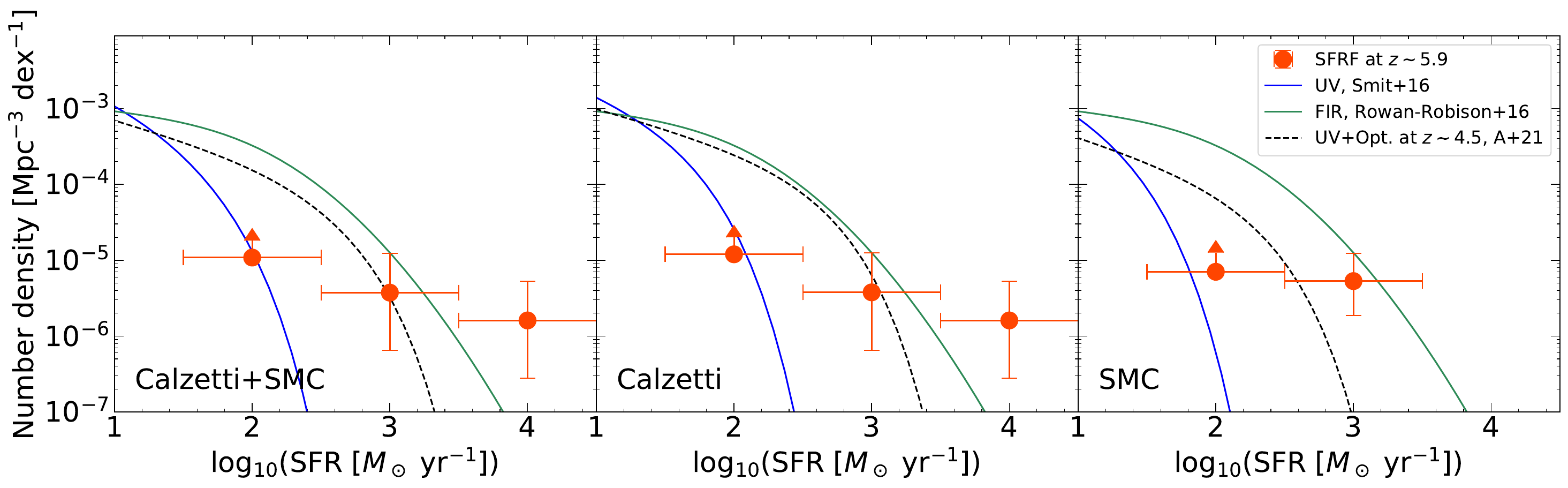}
\caption{The SFRF at $z\sim5.8$ obtained by SED fitting in this work (red circles).
Left, middle, and right panel shows the SFRF using the results with Calzetti+SMC, Calzetti, and SMC estimation, respectively.
The vertical error bars are $1\sigma$ Poisson uncertainties by \citet{gehrels_confidence_1986}.
The blue and green solid line show SFRFs estimated with rest-frame UV and FIR observations at $z\sim6$, respectively.
The black dashed line shows the SFRF at $z\sim4.5$ derived in a similar way to this work.
}
\label{fig:SFRF}
\end{figure*}

In Figure \ref{fig:SFRF}, we show the resulting SFRFs obtained under the each assumption of dust attenuation law (Calzetti+SMC, Calzetti, and SMC estimation).
The uncertainty of the SFRF is calculated assuming the Poisson uncertainty by \citet{gehrels_confidence_1986}.
As discussed in Section \ref{subsec:method_sfrf}, the completeness limit of the sample is $\sim10^{2.5}\ M_\odot\ {\rm yr}^{-1}$, thus the data point below that value is treated as a lower limit.
In the figure, we also show several SFRFs from literature for comparison converted to the same IMF as we use.
\citet[S16 hereafter]{smit_inferred_2016} derived an SFRF based on the rest-frame UV LF at $z\sim6$ after corrected for the dust attenuation with assuming the $A_{\rm UV}$-$\beta$ relation.
S16 assumed \citet{meurer_dust_1999}- and SMC-type $A_{\rm UV}$-$\beta$ relation and derived the respective SFRFs.
For a fair comparison, we compare our result to SFRFs in the literature under the corresponding assumption of dust attenuation law.
For result using Calzetti+SMC estimation (left panel in Figure \ref{fig:SFRF}), we simply compare to the average of the SFRFs with Calzetti and SMC law from the literature.
In addition, we show the SFRF obtained from FIR observations at $z\sim6$ estimated by \citet{rowan-robinson_star_2016} in the figure.
We have to keep it in mind that this SFRF from FIR observations was measured only in the brightest region (${\rm SFR}\sim10^4\ M_\odot\ {\rm yr}^{-1}$) and obtained by extrapolating down to the fainter region.

The SFRF obtained in this work (red circles) clearly shows an excess compared to that estimated from dust-corrected rest-frame UV LF (blue solid line).
Although the data points of the SFRF above ${\rm SFR}=10^{2.5}\ M_\odot\ {\rm yr}^{-1}$ contain a small number of galaxies (2 galaxies) and the uncertainties on these data points are large, the lower limit at the bin of ${\rm SFR}=10^{1.5-2.5}\ M_\odot\ {\rm yr}^{-1}$ also suggests that the SFRF obtained in this work is in an excess compared to the rest-frame UV-based estimation, and thus the excess seems to be rather robust.
This result is similar to that by A21, which showed the SFRF at $z\sim4.5$ derived from SED fitting shows an excess compared to that estimated from the dust-corrected rest-frame UV LF.
On the contrary, the resulting SFRF is roughly consistent with the FIR estimation given by \citet{rowan-robinson_star_2016} in spite of their large extrapolation.

\begin{figure}[tpb]
\centering
\plotone{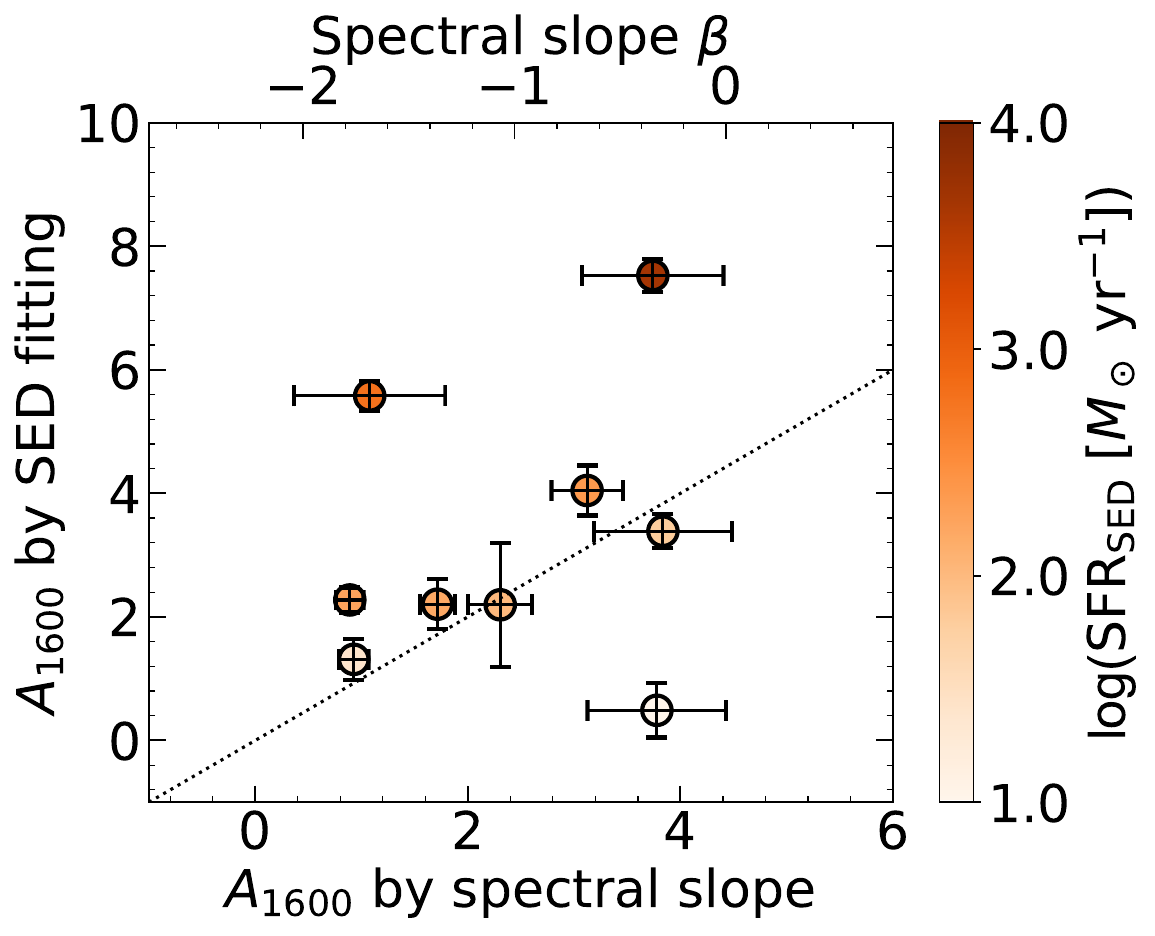}
\caption{The comparison of estimated amount of dust attenuation $A_{1600}$ from this work and from rest-frame UV-based approach which uses the $A_{\rm UV}$-$\beta$ relation (Equation \ref{eqn:M99}).
In this figure, we assume Calzetti attenuation law and use results with Calzetti estimation.
}
\label{fig:A1600_comparison}
\end{figure}

As discussed by A21, the difference in the SFRFs between that obtained with SED fitting on data from rest UV to optical and that estimated from the rest-frame UV LF stems from the difference in the estimation of the dust attenuation.
In Figure \ref{fig:A1600_comparison}, we compare the estimated amount of dust attenuation $A_{1600}$ in this work and that from the $A_{\rm UV}$-$\beta$ relation.
In the figure, we show the comparison with assuming Calzetti attenuation law, and use the $A_{\rm UV}$-$\beta$ relation given by \citet{meurer_dust_1999}:
\begin{equation}\label{eqn:M99}
    A_{1600} = 4.43 + 1.99\beta
\end{equation}
The rest-frame UV spectral slope $\beta$ is measured by fitting a power law to the rest-frame UV photometry for each object in the final sample.
The figure shows galaxies with larger SFRs tend to suffer from heavier dust attenuation, and the amount of dust attenuation on such galaxies is likely to be underestimated with the $A_{\rm UV}$-$\beta$ relation.
These results suggest that (1) previous approach to estimate the SFRF based only on the rest-frame UV observations underestimate the contribution from dust-obscured SF particularly in the brighter (larger SFR) region and (2) the contribution from dust-obscured SF can be well corrected for by using not only rest-frame UV but also optical data.

One may think the difference stems from a difference in sample selection.
Although we use catalogs selected in the rest UV wavelength (i.e., F160W detected), we make a sample based on photometric redshifts which is not exactly the same way as in part of rest-frame UV based works that uses Lyman break selection.
Practically, the Lyman break selection requests relatively blue in the rest-frame UV slope, while photometric redshift selection does not necessarily require.
In particular, two galaxies are mainly contributing to the highest-SFR bin where the excess is most significant, and these two galaxies can be the key origin of the excess.
However, both of the two galaxies meet the rest-frame UV slope criteria in $i_{814}$-dropout selection \citep[e.g.,][]{atek_are_2015} and their color can be classified as LBGs at this redshift.
Thus, the difference in sample selection is not expected to be the main reason for the difference in the estimated SFRFs.

\section{Implications from the SFRF}\label{sec:implication}
\subsection{Redshift evolution of the SFRF}\label{subsec:SFRFevo}
To see the redshift evolution of the SFRF, we also compare the result with the SFRF at $z\sim4.5$ derived from SED fitting by A21 (black dashed lines in Figure \ref{fig:SFRF}).
The number density derived in this work at the high-SFR end region is not largely different from that at $z\sim4.5$.
This suggests that the SFRF obtained with an approach using data from rest-frame UV to optical does not significantly evolve from $z\sim4.5$ to $z\sim5.9$ at the high-SFR end region.
In this subsection, we will further investigate the redshift evolution of the SFRF obtained with this approach.

One way to quantify the evolution of the SFRF is to parameterize the SFRF with an analytical form such as Schechter function.
Thus, we first fit the Schechter function to the SFRF derived in this work.
The Schechter function can be written as
\begin{align}
\phi(\dot{M}_\star) d(\log \dot{M}_\star) & \notag\\
= (\ln10)\phi^\star &\left( \frac{\dot{M}_\star}{\rm SFR^*}\right)^{\alpha+1} \exp\left[-\frac{\dot{M}_\star}{\rm SFR^*} \right]d(\log \dot{M}_\star)
\end{align}
where ${\rm SFR}^*$, $\phi^*$, and $\alpha$ is the characteristic SFR, the characteristic number density, and the low-SFR end slope, respectively.
Since the SFRF obtained in this work puts a constraint only in the high-SFR end region and the number of data points is small, we fix the $\alpha$ and $\phi^*$ and allow only ${\rm SFR}^*$ to vary.
We fix $\alpha$ to the same as that of the SFRF from rest-frame UV LF at $z\sim6$ (S16) since only rest-frame UV observations give constraints on $\alpha$ at this redshift: $\alpha=-1.63$ and $-1.72$ for Calzetti and SMC, respectively.
We also fix $\phi^*$ to the same as that estimated in a similar approach at $z\sim4.5$ (A21): $\phi^* = 7.24\times10^{-5}$ and $3.89\times10^{-5}$ for Calzetti and SMC, respectively.
We search for the best-fit parameter of ${\rm SFR}^*$ by minimizing $\chi^2$, and the lower limit on the SFRF at the bin of ${\rm SFR}=10^{2.0}\ M_\odot\ {\rm yr}^{-1}$ is also taken into account in calculating $\chi^2$ by reference to the approach given by \citet{sawicki_sedfit_2012}:
\begin{align}
\chi^2 = \sum_i \left(\frac{\phi_{{\rm d},i} - \phi_{{\rm m},i}}{\sigma_i} \right)^2 & \notag\\
- 2\sum_j\ln\int_{\phi_{{\rm lim},j}}^{\infty} \exp&\left[-\frac{1}{2}\left( \frac{\phi - \phi_{{\rm m},j}}{\sigma_j} \right)^2 \right] d\phi \label{eqn:chisq_lolim}
\end{align}
Here, subscript $i$ and $j$ denotes bins where the observed SFRF is complete and not complete (lower limit), respectively.
$\phi_{{\rm d},i}$ is the observed SFRF at the $i$th bin, $\sigma_i$ is its uncertainty, 
$\phi_{{\rm lim},j}$ is the lower limit on the SFRF at the $j$th bin, and $\phi_{{\rm m},i}$ or $\phi_{{\rm m},j}$ is the SFRF calculated from the Schechter function at the corresponding bin.
In this calculation using Equation (\ref{eqn:chisq_lolim}), we treat the lower limit at the bin of ${\rm SFR}=10^{2.0}\ M_\odot\ {\rm yr}^{-1}$ as 1$\sigma$ lower limit.

\begin{deluxetable}{ccc}
\tablenum{2}\label{tab:best_SFR*}
\tablecaption{The Best-fit Characteristic SFR and the Lower Limit on the SFRD}
\tablewidth{0pt}
\tablehead{
\colhead{Dust attenuation} & \colhead{$\log_{10}({\rm SFR}^*)$} & \colhead{$\log_{10}\rho^{\rm lim}_{\rm SFR}$} \\
\colhead{} & \colhead{$(M_\odot\ {\rm yr}^{-1})$} & \colhead{$(M_\odot\ {\rm yr^{-1}\ Mpc^{-3}})$}
}
\decimalcolnumbers
\startdata
Calzetti+SMC & $2.51^{+0.15}_{-0.87}$ & $-1.96$  \\
Calzetti & $2.36^{+0.12}_{-0.90}$ & $-1.96$ \\
SMC & $2.66^{+0.09}_{-0.42}$ & $-2.14$
\enddata
\end{deluxetable}

The resulting best-fit ${\rm SFR}^*$ is shown in Table \ref{tab:best_SFR*}.
Regardless of the assumption on the dust attenuation law, the best-fit ${\rm SFR}^*$ does not change significantly and is consistent with each other within the uncertainties.
Since recent ALMA observations in the high-$z$ universe suggest SMC-like attenuation law can be more appropriate in modeling high-$z$ SF galaxies \citep[e.g.,][]{fudamoto_alpine-alma_2020,schouws_significant_2021}, we use the result with SMC estimation to discuss the evolution of the SFRF hereafter in this subsection.

To see the redshift evolution, we next compile measurements of the SFRF at a wide range of redshift using various probes of the SFR, and compare the best-fit Schechter parameters.
At $z\lesssim3$, investigations with various probes are available and we compile results using H$\alpha$ \citep{sobral_large_2013}, infrared \citep{magnelli_evolution_2011}, and bolometric \citep{bell_star_2007,reddy_multiwavelength_2008} observations.
At $z\gtrsim3$, on the other hand, we compare results using (dust-corrected) rest-frame UV LFs (S16) and those with not only rest-frame UV but also optical data including this work (A21; A22).
For a fair comparison, from S16, we use results using an SMC-type $A_{\rm UV}$-$\beta$ relation in the dust correction of the UV LFs:
\begin{equation}
    A_{1600} = 2.45 + 1.1\beta
\end{equation}
Results in the literature using other IMF are converted to the same IMF as we use in this work.
When best-fit Schechter parameters for the SFRF are not quoted in literature, we derive the best-fit parameters by ourselves using the result in the literature.
The detailed procedure for this fit is given in Appendix \ref{apd:Schechter_fit}.

\begin{figure}[tpb]
\centering
\includegraphics[width=1.0\columnwidth, angle=0]{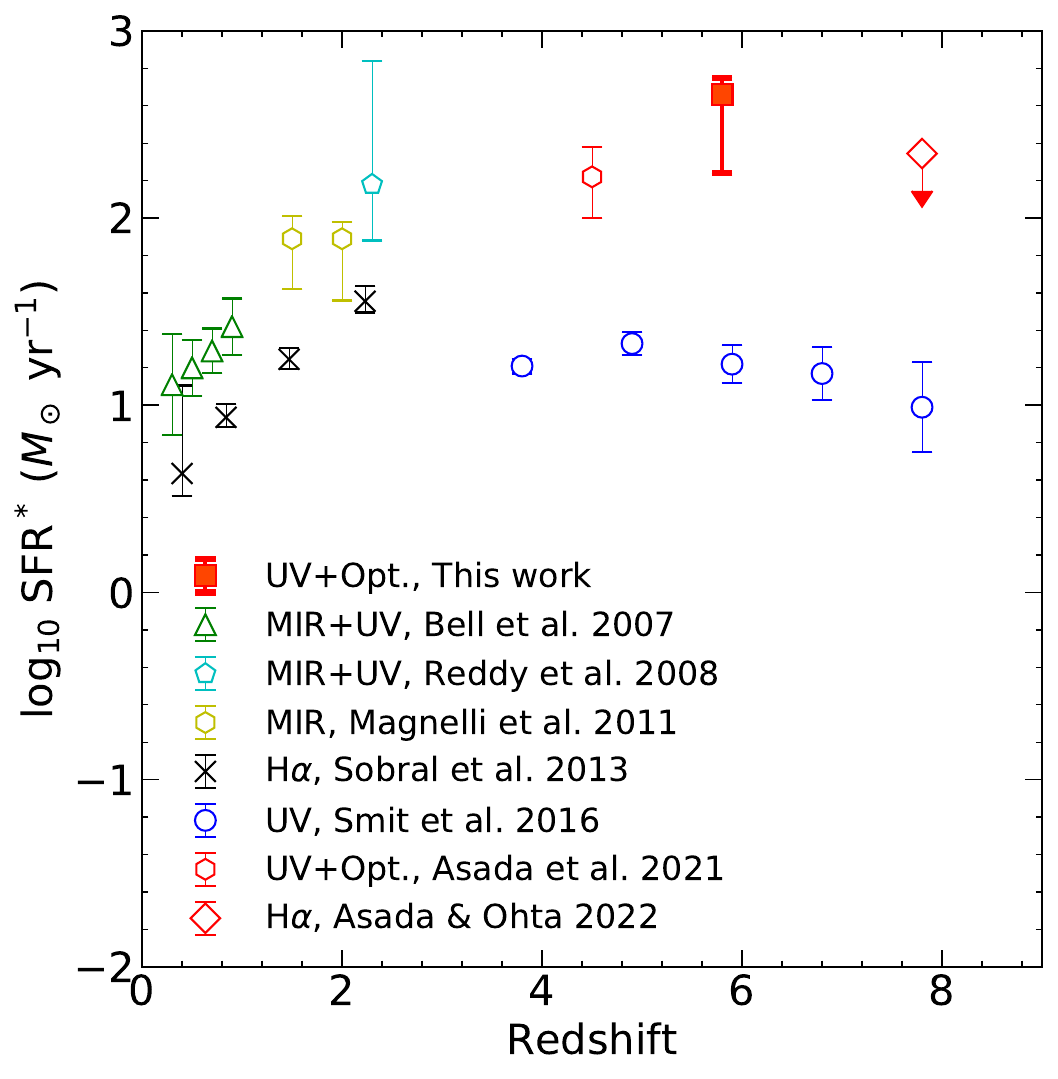}
\caption{The redshift evolution of the characteristic SFR ${\rm SFR}^*$.
At $z\gtrsim3$, investigations based only on rest-frame UV observations are shown by blue symbols, and those using rest-frame UV to optical observations are shown by red symbols including the result in this work.
}
\label{fig:SFRFparam_evo}
\end{figure}

Figure \ref{fig:SFRFparam_evo} shows the resulting evolution of the ${\rm SFR}^*$.
At $z\lesssim3$, results with various probes are roughly consistent with each other, although the estimations from H$\alpha$ LF give slightly lower values.
As discussed by \citet{smit_star_2012}, this may be originated from an incomplete sampling or inappropriate correction for the dust attenuation of the H$\alpha$ LFs.
The ${\rm SFR}^*$ decreases from $z\sim2$ to the local by $\sim1$ dex, which means the star formation activities in galaxies are typically suppressed from $z\sim2$ to the present universe.
Recent hydrodynamical cosmological simulation claimed that the main mechanism of this quenching is AGN feedback \citep[e.g.,][]{katsianis_evolution_2017}.
Other observational studies \citep[][]{harikane_goldrush_2022} claimed this quenching is attributed to a decrease of the dark-matter accretion rate onto halos due to the cosmic expansion.

At $z\gtrsim3$, the ${\rm SFR}^*$s estimated with the approach using rest-frame UV to optical data (red symbols in Figure \ref{fig:SFRFparam_evo}) show as high values as that at $z\sim2$, while those from rest-frame UV LFs (blue symbols in Figure \ref{fig:SFRFparam_evo}) show lower values than at $z\sim2$ by $\sim1$ dex throughout $z\gtrsim3$.
With the previous measurements using rest-frame UV LFs, ${\rm SFR}^*$ is thought to decrease from $z\sim2$ to $z\sim8$ \citep[e.g.,][]{smit_star_2012}.
However, the SFRF measurements using rest-frame optical data indicate the ${\rm SFR}^*$ is roughly constant from $z\sim2$ to $z\sim6$ and may decrease above $z\sim6$.
Since the parameter ${\rm SFR}^\star$ is sensitive to the high-SFR end of the SFRF, this evolution of the ${\rm SFR}^\star$ suggests that the high-SFR end of the SFRF grows as early as by $z\sim6$, or in the epoch of reionization, to a similar value as at $z\sim2$ when the star formation activity in the universe reaches its peak.

\subsection{Star Formation Rate Density}\label{subsec:SFRD}
The star formation rate density (SFRD) is usually measured by integrating the SFRF down to a lower bound.
The rest-frame UV absolute magnitude of $M_{\rm UV}=-17.0$ mag is commonly adopted as the lower bound, which corresponds to a SFR of $\sim0.22\ M_\odot\ {\rm yr}^{-1}$.
The SFRF measurement in this work is only in much larger SFR region than this lower bound of integration.
We here calculate the SFRD in this work as follows:

\begin{equation}
    \rho_{\rm SFR}^{\rm lim} = \sum_i^{N_{\rm gal}} \frac{1}{V^{\rm eff}_i} \int \dot{M}_\star P_i(\dot{M}_\star) d\dot{M}_\star
\end{equation}
where the notations are the same as those in Equation (\ref{eqn:SFRF_calc}).
Since the summation is taken for galaxies in the final sample which is complete only above the completeness limit (${\rm SFR}=10^{2.5}\ M_\odot\ {\rm yr}^{-1}$), this values gives a lower limit on the SFRD.
We compare this limit with the previous measurements using the lower bound of integration, $M_{\rm UV}=-17.0$ mag.

\begin{figure*}[tpb]
\centering
\includegraphics[width=2.0\columnwidth, angle=0]{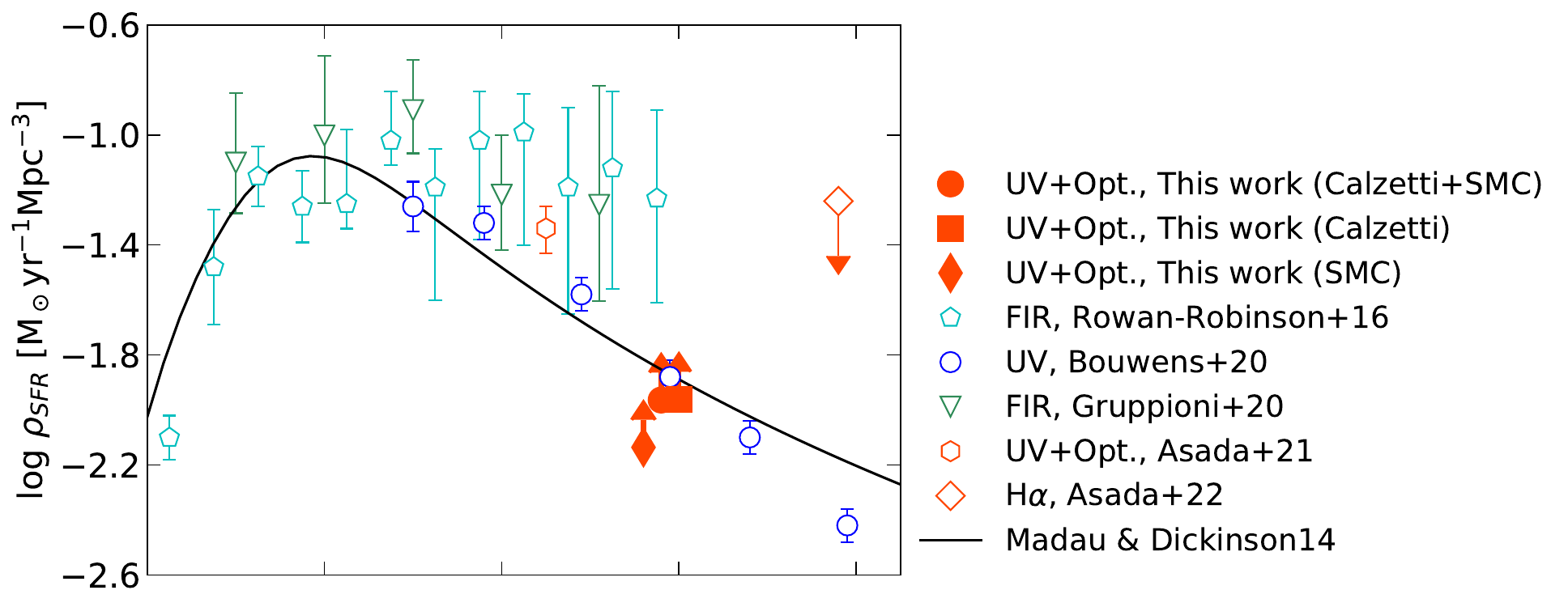}
\caption{The SFRD estimation and comparisons with the literature.
The lower limits obtained in this work using results by Calzetti+SMC (red filled circle), Calzetti (red filled square), and SMC (red filled diamond) estimation are shown.
Recent SFRD estimations using various probes are also shown for comparison.
For clarity, our results with Calzetti (SMC) law is shifted by $\Delta z=0.2$ ($-0.2$), and the result at $z=5.25$ by \citet{gruppioni_alpine-alma_2020} is shifted by $\Delta z=-0.15$.
}
\label{fig:SFRD}
\end{figure*}

The results are shown in Figure \ref{fig:SFRD} and Table \ref{tab:best_SFR*}.
In the figure, for a comparison, we also show recent measurements of SFRD from literature using FIR \citep{rowan-robinson_star_2016,gruppioni_alpine-alma_2020}, rest-frame UV \citep{bouwens_alma_2020}, both rest-frame UV and optical (A21), and H$\alpha$ (A22) observations.
At $z\gtrsim5$, part of FIR observations \citep[e.g.,][]{rowan-robinson_star_2016,gruppioni_alpine-alma_2020} indicate a more intensive star formation activities than dust-corrected rest-frame UV observations \citep[e.g.,][and see Figure \ref{fig:SFRD}]{madau_cosmic_2014,bouwens_alma_2020}.
With Calzetti+SMC (the red filled circle) or Calzetti (the red filled square) estimation, the lower limit obtained in this work suggests that the estimations from rest-frame UV observation are marginally acceptable or underestimate the SFRD, and the measurements with FIR observations are more likely, even though the lower limit does not include the contribution from fainter (lower SFR) galaxies.
On the contrary, the constraint obtained with SMC estimation (the red filled diamond) is consistent with both the measurements with FIR and rest-frame UV observations and not tight enough to determine which measurements are more likely.

It is worth noting that the assumption on the dust attenuation law can affect the SFRD estimations.
Particularly, if SMC law is assumed instead of Calzetti law, the estimated SFRD can decrease.
\citet{madau_cosmic_2014} (black solid line in the figure) calculated the correction factor for dust attenuation using Equation (\ref{eqn:M99}), which is almost the same as a correction using Calzetti law.
From A21, we show the estimation using Calzetti law in Figure \ref{fig:SFRD} (the open red hexagon), but A21 discuss that the SFRD can be decreased by 0.2-0.6 dex when SMC law is assumed instead.
\citet{bouwens_alma_2020} (open blue circles) assumed a Calzetti-like and SMC-like $A_{\rm UV}$-$\beta$ relation for high-mass and low-mass galaxies, respectively.
Thus, the comparison of our result especially with SMC law to the literature may not be fair in that different assumptions are adopted.
Nevertheless, the consequences from comparisons with previous measurements using several probes above do not change.
At $z\sim6$, the dust correction calculated from rest-frame UV observations (i.e., $A_{\rm UV}$-$\beta$ relation) is $\sim0.20$ dex \citep{bouwens_alma_2020} and thus the dust-uncorrected SFRD is smaller than the dust-corrected value shown in the figure by $\sim0.20$ dex.
Even with the dust-uncorrected value, the lower limit obtained in this work with the SMC estimations is still consistent, which is the same consequence as above.
For comparisons with Calzetti+SMC and Calzetti estimation, the results are strengthened if the estimations with rest-frame UV observations are decreased.

\section{Discussions}\label{sec:discussion}
\subsection{Comparison of other physical properties to the previous studies}\label{subsec:Other}
In Section \ref{sec:sfrf}, we derived an SFRF based on SED fitting, and showed the SFRF is in an excess compared to that estimated only from rest-frame UV observations.
This difference may be originated in a difference of the sample (e.g., field-to-field variations) or a systematic difference in the SED modelings in this work.
To examine these possibilities, first, we derive the rest-frame UV LFs using the sample of galaxies in this work and compare them with those in the previous works.
We derive the rest-frame UV LFs with both the phot-$z$ sample and final sample.
The (dust-uncorrected) rest-frame UV absolute magnitude for each galaxy is calculated using the photometric redshift $z_{\rm best}$ and the observed flux in the rest-frame UV wavelength.
With the final sample, we derive the UV LF using the similar approach as we used in deriving the SFRF (Equation (\ref{eqn:SFRF_calc})).
With the phot-$z$ sample, since this sample is not subject to selections in IRAC photometry (i.e., the S/N cut in 5.8- and 8.0-$\mu$m band and the elimination of blended source in IRAC images), the UV LF is derived without the corresponding incompleteness correction factors: $f^{\rm sel}$ and $f_k^{\rm ch3}$ in Equation (\ref{eqn:eff_vol}).

\begin{figure}[tpb]
\centering
\plotone{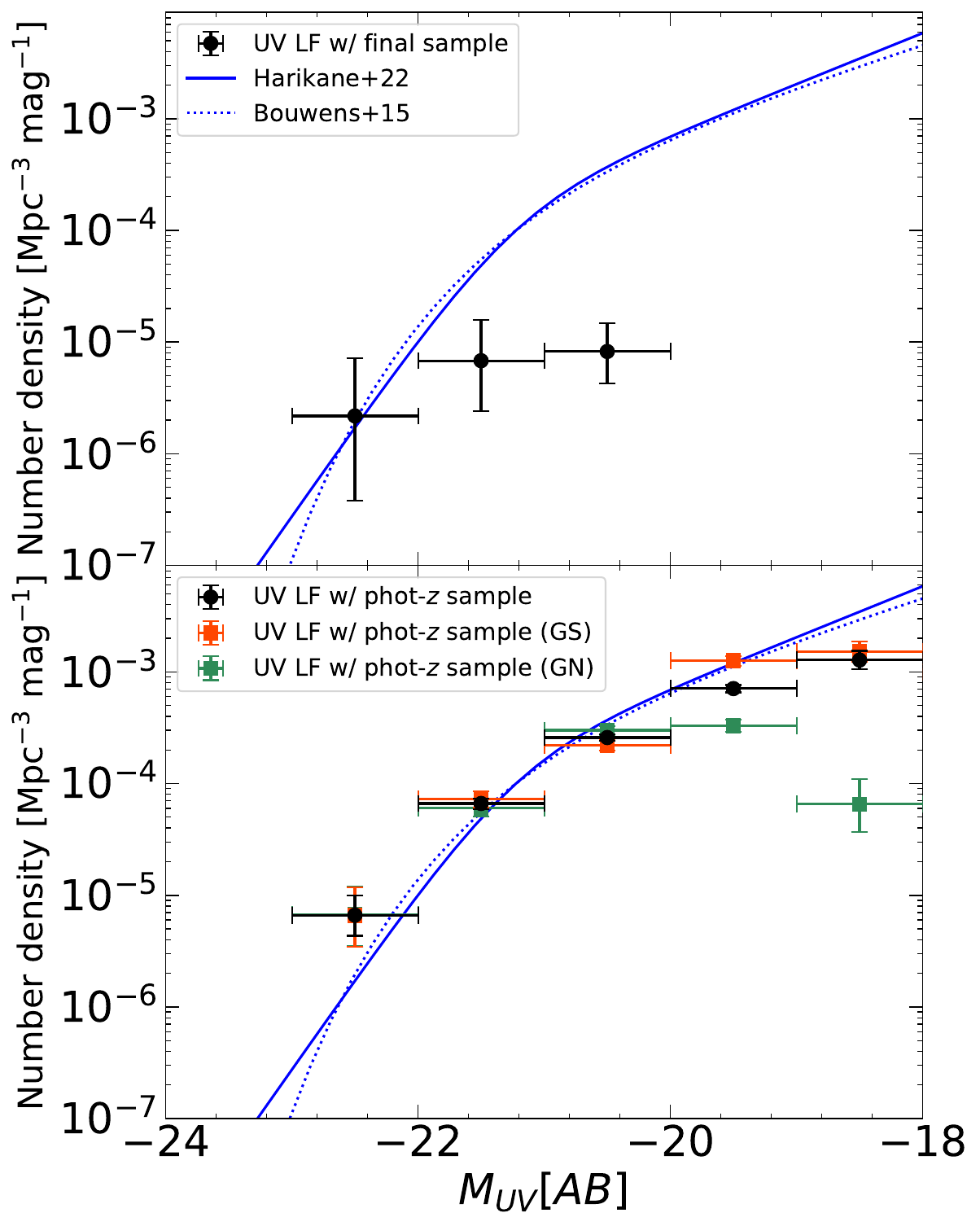}
\caption{The (dust-uncorrected) rest-frame UV LFs derived with the final (top) and phot-$z$ (bottom) sample (black circles).
The solid and dotted blue lines show the best-fit UV LF by \citet{harikane_goldrush_2021} and \citet{bouwens_uv_2015}, respectively.
In the bottom panel, we also show the UV LFs with galaxies in the phot-$z$ sample in GOODS-South (North) field by red (green) squares.
}
\label{fig:UVLF}
\end{figure}

The resulting UV LFs with the final and phot-$z$ sample are shown in the top and bottom panel of Figure \ref{fig:UVLF}, respectively.
For comparison, best-fit UV LFs in literature are also shown \citep{bouwens_uv_2015,harikane_goldrush_2021}.
The SFRF derived by S16 (blue solid line in Figure \ref{fig:SFRF}) is based on the UV LF by \citet{bouwens_uv_2015}, and thus we show it for comparison (blue dotted line in Figure \ref{fig:UVLF}).
\citet{harikane_goldrush_2021} found a significant excess of the rest-frame UV LFs compared to the previous best fit UV LFs written in the analytical form of Schechter function in the bright-end region, and suggested that the rest UV LF is well fitted by a double power law (DPL) rather than Schechter function.
To see if the excess in the SFRF found in this work stems from this excess in the UV LFs, we also show the UV LF by \citet{harikane_goldrush_2021} in the figure (blue solid line).

The UV LFs obtained with both the phot-$z$ sample and final sample agree well with those in the literature.
In particular, the difference between DPL and Schechter function fit (blue solid and dotted lines) emerges at the brighter population of galaxies ($M_{\rm UV} < -23$ mag) than that is used in this work.
This indicates that the excess of the SFRF is not caused by the difference of the (dust-uncorrected) rest UV LF itself but by the amount of dust correction.

In addition, we derive the UV LFs with the phot-$z$ sample in GOODS-S (red squares) and GOODS-N (green squares) fields.
The resulting UV LFs are in a good agreement with each other and with that of the phot-$z$ sample as a whole (black circles).
This indicates that there is no field-to-field variance between GOODS-S and -N, and our approach to derive the statistical properties such as SFRF and UV LF with combining sample of galaxies in GOODS-S and -N works well.
Although the UV LF with GOODS-N phot-$z$ sample seems to be systematically underestimated at the fain-end region ($M_{\rm UV}\sim-19$ mag), this is simply because the GOODS-N sample is incomplete at this region while the GOODS-S sample (and thus the phot-$z$ sample as a whole) is still complete.
The max-depth of the rest-frame UV observation by HST is different in the GOODS-S and -N due to the presence of Hubble Ultra Deep Field region in the GOODS-S field (see the top panel in Figure \ref{fig:mags}), and thus the completeness limit of the UV LF for GOODS-N sample is shallower than those for GOODS-S and phot-$z$ sample.

Next, to examine if there is a systematic difference of SED modelings in this work and those in the previous works, we derive the galaxy stellar mass function (GSMF) using the result of SED fitting obtained in Section \ref{subsec:physpara}, and compare it with other studies.
Again, we use the same approach to derive the GSMF as we used to derive the SFRF (Equation (\ref{eqn:SFRF_calc})), and derive the GSMF under the each assumption of dust attenuation law (Calzetti+SMC, Calzetti, and SMC).
However, the difference in the resulting GSMFs with the different attenuation laws is much smaller than that in the SFRFs and the three GSMFs are almost identical.
Thus, we only use the resulting GSMF with the Calzetti estimation hereafter.

\begin{figure}[tpb]
\centering
\plotone{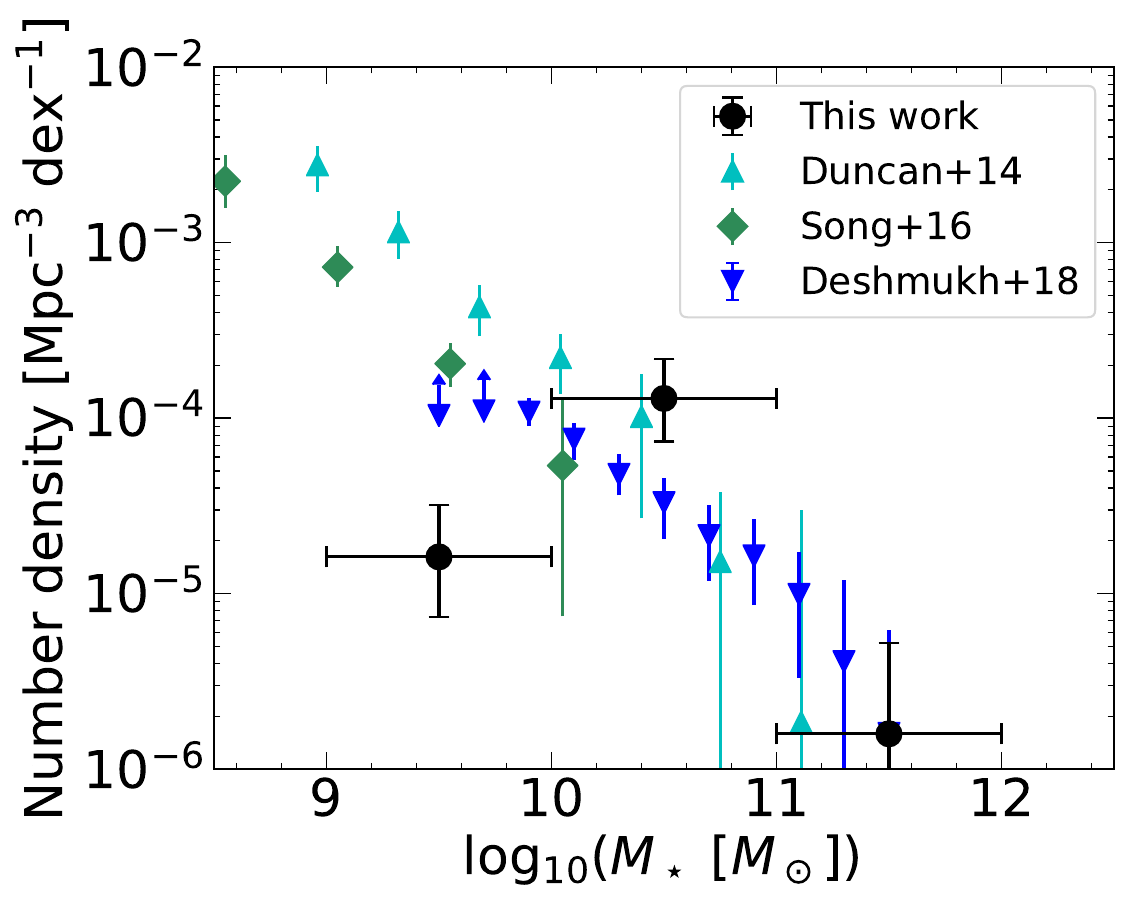}
\caption{The GSMF measurement at $z\sim6$ and comparison to the previous works.
In this figure, only the GSMF obtained using the result of Calzetti estimations is shown, but the difference of dust attenuation law has negligible effects on the resulting GSMF.
}
\label{fig:GSMF}
\end{figure}

The result is shown in Figure \ref{fig:GSMF}.
In the figure, the previous measurements of GSMF at $z\sim6$ are also shown for comparison \citep{duncan_mass_2014,song_evolution_2016,deshmukh_spitzer_2018}.
The resulting GSMF is in a good agreement with the previous estimations of the GSMF from the literature except for the lowest stellar mass bin ($M_\star = 10^{9.5}\ M_\odot$), where our sample is incomplete since the detection limit in 5.8$\mu$m (and 8.0$\mu$m) band in this work is shallow.
Since the GSMFs in the previous works are also derived based on SED fitting, significant difference in the SED modeling between this work and literature is not indicated from this comparison.

\subsection{Consistency with observations in other wavelengths range}
We have used photometry from rest-frame UV to optical and derived the SFRF and other physical properties.
On the contrary, part of galaxies in the final sample are also observed in other wavelengths range such as far-infrared (FIR) and radio.
In this subsection, we compare our result of SED fittings with the FIR and/or radio observation to see the consistency between the estimations with SED fitting and observations.

To compare with FIR observations, we calculate the total FIR luminosity ($L_{\rm TIR}$) from the result of SED fitting.
Namely, we calculate the total energy attenuated by dust for each of the best-fit model SEDs obtained in Section \ref{subsec:SEDfit} and \ref{subsec:two-comp}, and regard this energy to be re-emitted in FIR through the dust thermal emission assuming the energy conservation. 
Using the best-fitting $L_{\rm TIR}$, we derive three set of weighted means and weighted standard deviations of $\log_{10} (L_{\rm TIR})$ for each of galaxies in the final sample with the Calzetti, SMC, and Calzetti+SMC estimation as we did for other physical parameters such as the SFR and stellar mass in Section \ref{subsec:physpara}.
Here, we calculate the weighted mean of $\log_{10} (L_{\rm TIR})$ rather than $L_{\rm TIR}$ itself since the estimated total energy from SED fitting largely changes by several order of magnitudes with the model assumptions.
When the best-fitting $L_{\rm TIR}$ is 0 (i.e., a dust-free solution $E(B-V)=0$), its logarithm is treated as a sufficiently small number ($\log(L_{\rm TIR}/L_\sun) = 9.3$, which corresponds to ${\rm SFR_{IR}}\sim0.2 \ M_\odot\ {\rm yr}^{-1}$).
The resulting estimations are shown in Table \ref{tab:params_all}.

\subsubsection{FIR and/or radio observations}\label{subsubsec:FIRobs}
In this subsection, we introduce the observations in FIR and/or radio wavelengths available from literature.

\begin{itemize}
    \item 3073-GS
\end{itemize}
The galaxy 3073-GS is also selected as the target in the ALPINE-ALMA survey\footnote{The ID of this galaxy in ALPINE survey is CANDELS\_GOODSS\_14.} \citep{le_fevre_alpine-alma_2020}.
The galaxy is observed in ALMA Band 7 and its dust continuum emission is not detected with an upper limit of $\log_{10}(L_{\rm TIR}/L_\odot)<11.31$  \citep{bethermin_alpine-alma_2020}.

\begin{itemize}
    \item 17541-GS and 30423-GS
\end{itemize}
These two galaxies of 17541-GS and 30423-GS are located in the region covered with an unbiased survey in ALMA Band 6 by GOODS-ALMA survey \citep{franco_goods-alma_2018,franco_goods-alma_2020}.
There are no counterparts of 17541-GS or 30423-GS in the GOODS-ALMA catalog, and thus these two galaxies are not detected in ALMA Band 6.
The limiting flux of the catalog is $\sim640\ \mu$Jy.
Using an FIR SED template given by \citet{schreiber_dust_2018}, this corresponds to an upper limit of $\log_{10}(L_{\rm TIR}/L_\odot)<12.59$.

\begin{itemize}
    \item 29111-GN
\end{itemize}
The galaxy 29111-GN is detected in a wide wavelength range from MIR to radio.
The radio emission from the galaxy is detected at $1.4$ GHz with the Very Large Array (VLA) with a flux of $f_\nu = 25.4\pm3.2\ \mu$Jy \citep{owen_deep_2018}.
\citet[L18 hereafter]{liu_super-deblended_2018} used the photometry at Spitzer/MIPS 24 $\mu$m and VLA 1.4 GHz in GOODS-N field as the prior information to deblend the highly confused images by Herschel and ground-based FIR to (sub-)millimeter observations.
For this galaxy\footnote{The ID of this galaxy in L18 is 14914}, L18 obtained an SED from MIR to radio and estimate the total FIR luminosity of $\log_{10}(L_{\rm TIR}/L_\odot)=13.24\pm0.03$.

In addition to the total FIR luminosity, for the galaxy, we compare the SFR derived from SED fitting with that directly estimated from the radio observation.
Given the photometric redshift $z_{\rm ph}=5.67$, the radio emission observed at 1.4 GHz corresponds to the emission at a rest-frame frequency of 9.3 GHz, where the non-thermal (synchrotron) emissions from supernova remnants are dominant in bright starburst galaxies.
\citet{condon_radio_1992} obtained a conversion factor from (non-thermal) radio luminosity $L_{\rm NT}$ to the SFR for massive ($>5\ M_\odot$) stars assuming the extended Miller-Scalo IMF.
For a fair comparison, starting with the formulation given by \citet{condon_radio_1992} that links the supernova rate and (non-thermal) radio luminosity $L_{\rm NT}$, we calculate the conversion factor between $L_{\rm NT}$ and SFR for the mass range of 0.08-120 $M_\odot$ under the same assumption of IMF (i.e., Chabrier03 IMF):
\begin{equation}\label{eqn:radioSFR}
    \left( \frac{L_{\rm NT}}{\rm W\ Hz^{-1}} \right) \sim 1.5\times10^{21} \left( \frac{\nu_{\rm rest}}{\rm GHz} \right)^{-\alpha} \left( \frac{\rm SFR}{M_\odot\ {\rm yr}^{-1}} \right)
\end{equation}
where $\nu_{\rm rest}$ is the rest-frame frequency of the observation and $\alpha$ is the spectral index (see Appendix \ref{apd:SFR_rad} for the detail derivation of this conversion factor).
Here we use $\alpha=0.8$ \citep{condon_radio_1992} and obtain ${\rm SFR_{\rm rad}}=5286\pm666\ M_\odot\ {\rm yr}^{-1}$ from the VLA observation.

\subsubsection{Comparisons to FIR and/or radio observations}
\begin{figure}[tpb]
\centering
\includegraphics[width=0.9\columnwidth, angle=0]{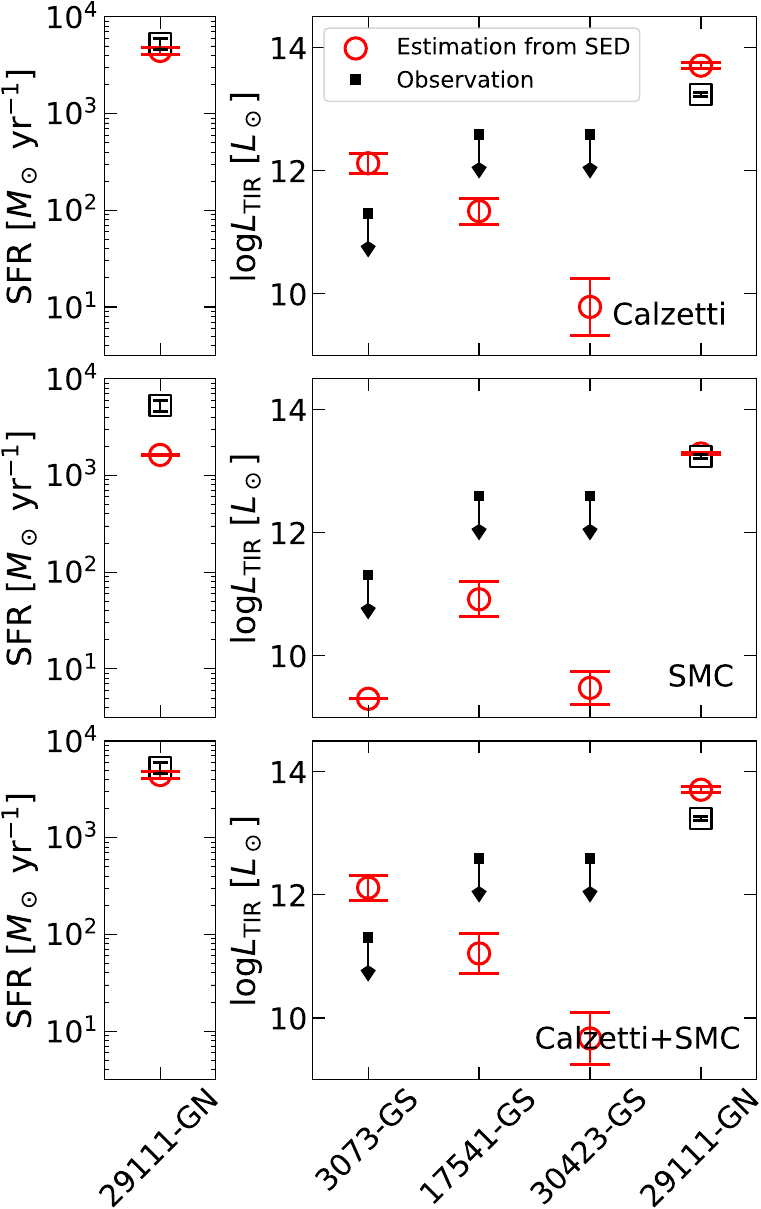}
\caption{Comparisons of the SFR (left panel) and $L_{\rm TIR}$ (right panel) derived from SED fitting using data from rest-frame UV to optical (red circles) with those derived from independent observations in other wavelengths range (black squares).
We show the comparison with Calzetti, SMC, and Calzetti+SMC estimations in the top, middle, and bottom panel, respectively.
As for the SFR, we compare the SFR estimated from SED fittings and that from the radio flux density using Equation (\ref{eqn:radioSFR}).
See text for details.}
\label{fig:SEDvsObs}
\end{figure}

We compare the SFR and $L_{\rm TIR}$ derived from SED fitting using data from rest-frame UV to optical with those indicated from independent observations in other wavelength described in Section \ref{subsubsec:FIRobs}.
The result of comparisons is shown in Figure \ref{fig:SEDvsObs}.
In the figure, estimations from SED fitting and constraints from independent observations are shown by red and black symbols, respectively.

When Calzetti law is assumed as the dust attenuation law in SED fitting (top panels), $L_{\rm TIR}$ seems to be overestimated.
This indicates that using Calzetti law with SED fitting overestimates the amount of dust attenuation, which results in a larger value of the intrinsic rest-frame UV luminosity, possibly overestimation of the SFRF as well.

On the contrary, when SMC law is assumed (middle panels), the total FIR luminosity predicted from SED fitting is in a good agreement with the observations (middle right panel), while the SFR predicted from SED fitting using SMC law is in tension with that indicated from the radio observation (middle left panel).
This agreement in the total FIR luminosity is consistent with the previous results from recent ALMA observations in the high-$z$ universe based on analyses of the IRX-$\beta$ relation \citep[e.g.,][]{fudamoto_alpine-alma_2020,schouws_significant_2021}.
It is worth noting that these previous approaches based on the IRX-$\beta$ relation use only rest-frame UV and FIR observations and assume a simplified spectrum of galaxies by a power-law function in the rest-frame UV wavelength range, while the comparison in this work is a consequence from rest-frame UV, optical, and FIR observations by assuming more realistic spectra of galaxies.
Thus, this comparison further supports these recent studies that claim the SMC-like law is better representation for high-z galaxies.

Lastly, with the Calzetti+SMC estimation (bottom panels), the SFR estimated from SED fitting and radio observation agree with each other (bottom left panel), while the total FIR luminosity is overestimated with SED fitting (bottom right panel).

We have to keep it in mind that, in calculating the ${\rm SFR_{radio}}$ using Equation (\ref{eqn:radioSFR}), we do not consider the contribution from thermal radio emissions but regard all the radio flux observed with VLA as the non-thermal emission.
The contribution of thermal emission to the total radio luminosity at $\nu_{\rm rest}=9.3$ GHz is roughly $\sim25\%$ in the local starburst galaxy M82 \citep{condon_radio_1992}.
Inferring from this example, the non-thermal radio luminosity $L_{\rm NT}$ may get smaller than the total radio luminosity at $\nu_{\rm rest}=9.3$ GHz observed with VLA by a factor of $\sim0.75$, which results in decreasing ${\rm SFR_{radio}}$ (black squares in left panels) by the same factor.
However, this displacement ($\sim0.1$ dex) affects the ${\rm SFR_{radio}}$ only slightly and changes the ${\rm SFR_{radio}}$ within its uncertainty, thus this effect is negligible.

Considering these results of comparisons, estimations using SMC law as the dust attenuation law in SED fitting is consistent with independent observations.
However, estimations using Calzetti law or Calzetti+SMC seem to be in tension, and their resulting SFRF (and SFRD) may be overestimated.
Since the comparison is only for part of the final sample of this work, a further investigation is needed to examine more robustly including much deeper FIR observations and enlarging the sample size.

\subsection{AGN contamination}
In this work, we eliminate AGNs by removing X-ray sources brighter than $\gtrsim1\times10^{43}\ {\rm erg\ s^{-1}}$ (c.f., sample selection in \S \ref{subsec:sample}), thus the lower-luminousity AGNs may contaminate our sample.
It should be useful to roughly estimate the effect of these AGNs that may be remained in the sample.
Considering typical SEDs of AGNs \citep[e.g.,][]{elvis_atlas_1994,koratkar_ultraviolet_1999,nemmen_spectral_2014}, the $H_{160}$-band magnitude of an AGN at $z\sim6$ with X-ray luminosity of $\sim1\times10^{43}\ {\rm erg\ s^{-1}}$ is $\sim27\ {\rm mag}$, thus AGNs fainter than this can be detected if they locate in the deepest survey region (c.f., see the upper panel in Figure \ref{fig:mags}).
The X-ray luminosity limit corresponds to a UV absolute magnitude limit of $M_{\rm UV}\sim-19$ mag.
This means that our sample after the removal of X-ray sources (phot-$z$ sample and final sample) is expected not to contain AGNs brighter than $M_{\rm UV}\sim-19$ mag, and therefore only the faintest bin in the bottom panel of Figure \ref{fig:UVLF} may be contaminated by low-luminousity AGNs.

Nevertheless, it is also meaningful to estimate the effect of AGNs using AGN LFs at $z\sim6$ without assuming the AGN SEDs, because the SED templates are derived from low-$z$ AGNs and the SED may be different for AGNs at $z\sim6$.
\citet{matsuoka_subaru_2018} derived an AGN LF at $z\sim6$ over the range of $M_{\rm UV}$ from $-30$ to $-22$ mag.
The number density of AGNs was estimated to be $\sim2\times10^{-8}\ {\rm Mpc}^{-3}\ {\rm mag}^{-1}$ even at their most abundant and faintest bins of $M_{\rm UV}=-22$ or $-23$ mag.
This number density is less than 1\% of that obtained for the sample in this work at the magnitude bin (Figure \ref{fig:UVLF}).
This magnitude bin of $M_{\rm UV}\sim-22$-$-23$ mag is the brightest bin for our sample, and the fractions of AGNs in the fainter bins where most galaxies in our sample are included are expected to be several orders of magnitude smaller than 1\%, if we suppose the extrapolated fainter part of the AGN LF (Figure 12 by \citet{matsuoka_subaru_2018}).
Thus, the effect of AGN contamination is considered to be negligible.

Indeed, we can estimate the expected number of AGNs in each magnitude bin in Figure \ref{fig:UVLF} by extrapolating the AGN LF by \citet{matsuoka_subaru_2018}.
We calculate the maximum effective volume for the search of contaminating AGNs by assuming incompleteness correction factors in Equation (\ref{eqn:eff_vol}) to be unity.
With this effective volume and the extrapolated AGN LF, the expected number of AGNs is $\sim0.02$ at the bin of $M_{\rm UV}=-22.5$, and $\sim0.07$ at $M_{\rm UV}=-17.5$.
Taking the summation of these expected numbers suggests the maximum expected number of AGNs contaminating the phot-$z$ sample is $\sim0.25$.
We have to note that this estimation does not consider the effect of X-ray luminosity cut.
The low-luminousity AGNs which may remain in the sample even after the X-ray luminosity cut should be included in the faint magnitude bins, and the effective volume for the search of these faint sources is smaller than that we used here.
Therefore, the expected number of $0.25$ obtained here is an upper limit, and the actual number must be much smaller than this.
By considering the sample size of the {\it phot-$z$ sample} ($\sim550$), the effect of AGNs on the resulting SFRF is expected to be negligible.

\subsection{Incompleteness of the lowest-SFR bin}
As discussed in Section \ref{subsec:SFRF} and in Figure \ref{fig:SFRF}, the SFRF obtained in this work shows an excess as compared to that from dust-corrected rest-frame UV LFs.
Because not only the highest SFR bin(s) whose uncertainty is large but also the lower limit at the bin of ${\rm SFR}=10^{1.5-2.5}\ M_\odot\ {\rm yr}^{-1}$ suggests the excess, we conclude the excess seems to be rather robust.
However, given that the incompleteness at the lowest-SFR bin of ${\rm SFR}=10^{1.5-2.5}\ M_\odot\ {\rm yr}^{-1}$ is based on the upper limit on rest-frame H$\alpha$ EW $<4000$\AA\ (c.f., Section \ref{subsec:method_sfrf}) and this maximum value is achievable only with an extreme stellar population such as Pop III stars, the incompleteness at this bin may be less reliable.

To see if the lowest-SFR bin is robustly incomplete, we examine how the SFR completeness limit changes with different assumptions of the maximum EW(H$\alpha$) value.
As shown in Section \ref{subsec:method_sfrf}, we use the relation between the 5.8$\mu$m band magnitude and SFR that can be obtained when an EW(H$\alpha$) value is assumed, and estimate the SFR completeness limit that corresponds to the 5.8$\mu$m band detection limit ($\sim24.5$ mag).
In Section \ref{subsec:method_sfrf}, we used the most extreme case of EW(H$\alpha$)$=4000$\AA\ as the upper limit to obtain the completeness limit of ${\rm SFR}\sim10^{2.5}\ M_\odot\ {\rm yr}^{-1}$.
Here, we examine how small the {\it maximum} EW(H$\alpha$) should be to make the lowest-SFR bin complete.
We obtain that the SFR completeness limit can be as small as ${\rm SFR}=10^{1.5}\ M_\odot\ {\rm yr}^{-1}$ and the lowest-SFR bin can be complete only when we assume the {\it maximum} EW(H$\alpha$) is $\sim500$\AA.
However, star-formging galaxies (SFGs) at this redshift are expected to have EW(H$\alpha$)$\sim500$\AA\ on average \citep[e.g.,][]{faisst_coherent_2016}, and non-negligible fraction of SFGs can have EW(H$\alpha$)$>500$\AA.
This indicates that the SFR bin cannot be complete, and thus the arguments in Section \ref{subsec:SFRF} should be rather robust.

\section{Summary}\label{sec:summary}
In the high-$z$ ($z\gtrsim4$) universe, the contribution of dust-obscured star formation to the total star formation activity is still unclear, and an investigation of the SFRF is desired in an independent way of the previous approaches based on rest-frame UV observations or FIR observations.
Very recently, investigations at $z\gtrsim4$ with an approach using not only rest-frame UV but also optical data including H$\alpha$ emission line have been reported.
In this work, we derived an SFRF at $z\sim5.8$ with SED fitting using data from rest-frame UV to optical wavelength of galaxies in the CANDELS GOODS-South and -North fields.
In deriving the SFRF, we extensively examined the assumptions on the SED modeling including various SFHs, dust attenuation law, and a two-component model.
We obtained three resulting SFRFs by assuming the Calzetti attenuation law, the SMC law, and a combination of Calzetti and SMC laws.
We also examined the redshift evolution of the SFRF and SFRD at $z\gtrsim4$ obtained with the independent approach using rest-frame optical emissions.
Our main results are as follows:
\renewcommand{\theenumi}{\arabic{enumi}}
\begin{enumerate}
    \item The resulting SFRFs at $z\sim5.8$ show an excess compared to those estimated from rest-frame UV observations but are roughly consistent with those estimated from FIR observations, although the SFRF from FIR observations is actually measured only at the brightest region and obtained with a large extrapolation down to the fainter region (Figure \ref{fig:SFRF}).
    The excess in the resulting SFRFs is originated from the difference of the estimated amount of the dust attenuation (Figure \ref{fig:A1600_comparison}).
    This suggests that the contribution from the dust-obscured intensively star-forming galaxies to the total star formation activities at $z\sim6$ is underestimated with the previous approach based only on the rest-frame UV observations (Section \ref{subsec:SFRF}).
    \item The SFRF is parameterized by assuming the Schechter functional form.
    The low-SFR end slope and characteristic number density are fixed and best-fit ${\rm SFR}^*$ is estimated (Table \ref{tab:best_SFR*}).
    Together with SFRFs at $z\gtrsim4$ in literature estimated with the approach using rest-frame optical emissions, the characteristic SFR (${\rm SFR}^\star$) is roughly constant at $z\sim2$ to $z\sim6$ and may decrease above $z\sim6$.
    Since the parameter ${\rm SFR}^\star$ is sensitive to the high-SFR end of the SFRF, this suggests an early growth of the high-SFR end of the SFRF in the epoch of reionization, and the high-SFR end at $z\sim4$-6 is almost comparable to that at $z\sim2$ when the star formation activities reaches its peak (Section \ref{subsec:SFRFevo} and Figure \ref{fig:SFRFparam_evo}).
    \item Although the SFRFs obtained in this work are complete only in the high SFR region (${\rm SFR}>10^{2.5}\ M_\odot\ {\rm yr}^{-1}$), a lower limit on the SFRD is obtained simply by taking the summation of SFRs.
    From the estimation by assuming Calzetti law or Calzetti+SMC estimation, the previous measurements of the SFRD at $z\sim6$ based on rest-frame UV observations are marginally consistent or violate the lower limit obtained in this work, while the measurements based on FIR observations are acceptable and more likely.
    On the contrary, from the estimation by assuming SMC law, the lower limit on the SFRD is not tight enough to determine if the SFRD is underestimated or not with the previous measurements using rest-frame UV observations (Section \ref{subsec:SFRD} and Figure \ref{fig:SFRD}).
\end{enumerate}

We thank the anonymous referee for useful comments and suggestions to improve this paper.
Y.A. is supported by a Research Fellowship for Young Scientists from the Japan Society of the Promotion of Science (JSPS).
K.O. is supported by JSPS KAKENHI Grant Number JP19K03928.
This work is based on observations obtained with the NASA/ESA Hubble Space Telescope, retrieved from the Mikulski Archive for Space Telescopes (MAST) at the Space Telescope Science Institute (STScI). STScI is operated by the Association of Universities for Research in Astronomy, Inc. under NASA contract NAS 5-26555.

\software{Astropy \citep[][]{the_astropy_collaboration_astropy_2018}, APLpy \citep[][]{2012ascl.soft08017R}}

\appendix
\section{Physical properties of sample galaxies}\label{apd:physical_prop}
In this appendix, we show the physical properties of all the galaxies in the final sample including their SEDs and the best estimations obtained in Section \ref{subsec:physpara}.

\begin{figure*}[tpb]
\centering
\includegraphics[width=2.0\columnwidth, angle=0]{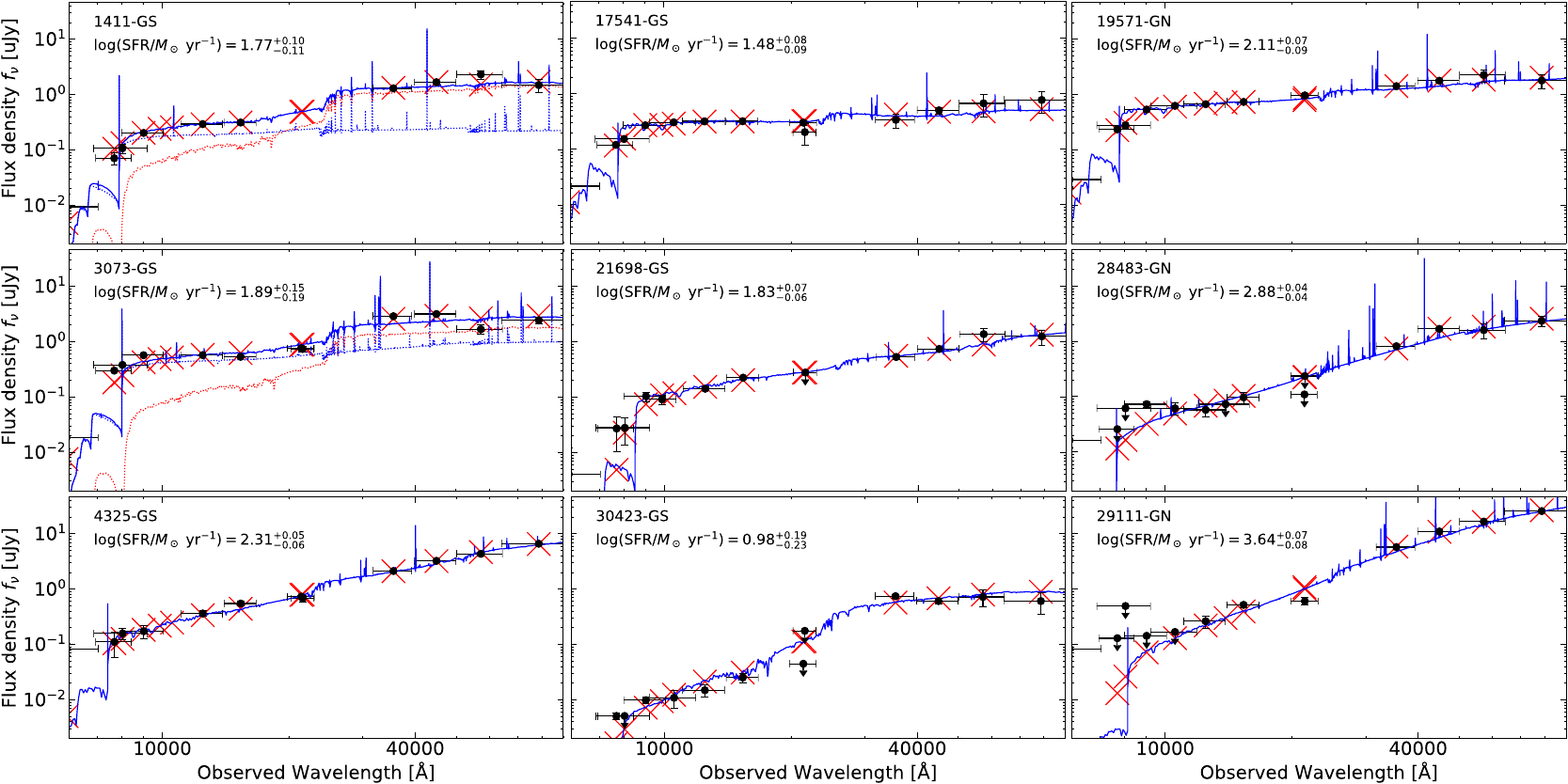}
\caption{SEDs of the galaxies in the final sample and best-fitting model SEDs.
In each panel, the black points with error bars represent the observed SED, the blue solid line shows the best fit model spectrum, and red crosses show the best fit model SED.
Among the 28 best-fitting models for each galaxy, in this figure, we only show the best-fitting model that gives minimum BIC (see also Section \ref{subsec:physpara}).
For galaxies that the two-component model give the minimum BIC, we also show the best-fitting spectrum of the old stellar population and young star-forming population with red and blue dotted lines, respectively.
At the top left in each panel, we also show the name of the galaxy and the SFR with Calzetti+SMC estimation.
}
\label{fig:SED}
\end{figure*}

Figure \ref{fig:SED} shows the SEDs of galaxies in the final sample.
We also show one of the best fit SEDs derived in Section \ref{subsec:SEDfit} and \ref{subsec:two-comp}.
Among the best-fitting SEDs, we only overplot the model SED that gives the minimum BIC.

In Table \ref{tab:params_all}, all the physical parameters obtained in this work are presented for each of the galaxies in the final sample with Calzetti, SMC, and Calzetti+SMC estimation.
In the physical parameter estimation, we also utilize the result of the two-component SED fitting (Section \ref{subsec:two-comp}).
As noted in Section \ref{subsec:two-comp}, the best-fit parameters of SFR, $M_\star$, and $L_{\rm TIR}$ in the two-component model are estimations for the whole system that includes both the old stellar population and the young star-forming population, but age and $A_V$ are estimations for the young star-forming population\footnote{In calculating $L_{\rm TIR}$, we only consider the contribution from the young star-forming population.
However, in this two-component model, we assume the old-stellar component is free from dust attenuation, and thus the dust thermal emission warmed up by the young star-forming population can be regarded as the total FIR emission of the galaxy.}.
Therefore, the best estimations for age and $A_V$, which are obtained using this two-component modeling, are not estimations for the whole system.
In particular, the best estimations of age obtained in this work do not denote time intervals since the onset of the first star formation episode that the galaxy experienced, but should be regarded as the time intervals since the onset of the last star formation episode.

\begin{deluxetable*}{clccccc}
\tablenum{3}\label{tab:params_all}
\tablecaption{The best estimations of physical parameters derived by SED fitting}
\tablewidth{0pt}
\tablehead{
\colhead{Dust attenuation} & \colhead{Name} & \colhead{Age [Myr]} & \colhead{$A_V$ [mag]} & \colhead{$\log (M_\star/M_\odot)$} &
\colhead{$\log ({\rm SFR}/M_\odot\ {\rm yr}^{-1})$} & \colhead{$\log (L_{\rm TIR}/L_\odot)$}
}
\decimalcolnumbers
\startdata
{} & 1411-GS & $176^{+229}_{-176}$ & $0.47^{+0.45}_{-0.45}$ & $10.44^{+0.09}_{-0.12}$ & $1.86^{+0.28}_{-1.01}$ &   $11.19^{+0.63}_{-0.63}$ \\
{} & 3073-GS & $12^{+60}_{-12}$ & $0.92^{+0.09}_{-0.09}$ & $10.55^{+0.03}_{-0.03}$ & $2.24^{+0.04}_{-0.04}$ &   $12.12^{+0.20}_{-0.20}$ \\
{} & 4325-GS & $183^{+132}_{-132}$ & $1.64^{+0.16}_{-0.16}$ & $10.71^{+0.07}_{-0.08}$ & $2.36^{+0.08}_{-0.10}$ &   $12.45^{+0.14}_{-0.14}$ \\
{} & 17541-GS & $77^{+65}_{-65}$ & $0.28^{+0.19}_{-0.19}$ & $9.35^{+0.18}_{-0.31}$ & $1.35^{+0.25}_{-0.70}$ &   $11.05^{+0.33}_{-0.33}$ \\
Calzetti+SMC & 21698-GS & $81^{+73}_{-73}$ & $0.99^{+0.37}_{-0.37}$ & $9.89^{+0.09}_{-0.11}$ & $1.67^{+0.14}_{-0.22}$ &   $11.81^{+0.24}_{-0.24}$ \\
{} & 30423-GS & $1000^{+0}_{-0}$ & $0.14^{+0.16}_{-0.16}$ & $10.33^{+0.05}_{-0.06}$ & $-1.29^{+0.37}_{-\infty}$ &   $9.67^{+0.42}_{-0.42}$ \\
{} & 19571-GN & $105^{+105}_{-105}$ & $0.88^{+0.20}_{-0.020}$ & $10.04^{+0.12}_{-0.17}$ & $2.16^{+0.15}_{-0.23}$ &   $12.11^{+0.25}_{-0.25}$ \\
{} & 28483-GN & $12^{+18}_{-12}$ & $2.27^{+0.10}_{-0.10}$ & $9.78^{+0.25}_{-0.69}$ & $2.80^{+0.08}_{-0.09}$ &   $12.63^{+0.09}_{-0.09}$ \\
{} & 29111-GN & $123^{+101}_{-101}$ & $3.06^{+0.11}_{-0.11}$ & $11.59^{+0.13}_{-0.18}$ & $3.65^{+0.03}_{-0.04}$ &   $13.71^{+0.05}_{-0.05}$  \\
\\
\hline
{} & 1411-GS & $309^{+226}_{-226}$ & $0.89^{+0.41}_{-0.41}$ & $10.38^{+0.13}_{-0.20}$ & $2.00^{+0.27}_{-0.91}$ &   $11.73^{+0.52}_{-0.52}$ \\
{} & 3073-GS & $11^{+55}_{-11}$ & $0.92^{+0.08}_{-0.08}$ & $10.55^{+0.03}_{-0.03}$ & $2.24^{+0.04}_{-0.04}$ &   $12.12^{+0.16}_{-0.16}$ \\
{} & 4325-GS & $183^{+132}_{-132}$ & $1.64^{+0.16}_{-0.16}$ & $10.71^{+0.07}_{-0.08}$ & $2.36^{+0.08}_{-0.10}$ &   $12.45^{+0.14}_{-0.14}$ \\
{} & 17541-GS & $56^{+37}_{-37}$ & $0.53^{+0.13}_{-0.13}$ & $9.44^{+0.08}_{-0.10}$ & $1.40^{+0.27}_{-0.89}$ &   $11.34^{+0.21}_{-0.21}$ \\
Calzetti & 21698-GS & $47^{+42}_{-42}$ & $1.38^{+0.11}_{-0.11}$ & $9.97^{+0.03}_{-0.03}$ & $1.79^{+0.05}_{-0.06}$ &   $12.04^{+0.11}_{-0.11}$ \\
{} & 30423-GS & $1000^{+0}_{-0}$ & $0.20^{+0.18}_{-0.18}$ & $10.35^{+0.05}_{-0.06}$ & $-1.30^{+0.36}_{-\infty}$ &   $9.79^{+0.46}_{-0.46}$ \\
{} & 19571-GN & $93^{+64}_{-64}$ & $0.90^{+0.16}_{-0.16}$ & $10.03^{+0.11}_{-0.15}$ & $2.17^{+0.14}_{-0.21}$ &   $12.13^{+0.20}_{-0.20}$ \\
{} & 28483-GN & $13^{+18}_{-18}$ & $2.27^{+0.10}_{-0.10}$ & $9.76^{+0.24}_{-0.60}$ & $2.79^{+0.06}_{-0.07}$ &   $12.63^{+0.07}_{-0.07}$ \\
{} & 29111-GN & $123^{+101}_{-101}$ & $3.06^{+0.11}_{-0.11}$ & $11.59^{+0.13}_{-0.18}$ & $3.65^{+0.03}_{-0.04}$ &   $13.71^{+0.05}_{-0.05}$  \\
\\
\hline
{} & 1411-GS & $73^{+172}_{-73}$ & $0.15^{+0.05}_{-0.05}$ & $10.48^{+0.02}_{-0.02}$ & $1.72^{+0.18}_{-0.31}$ &   $10.77^{+0.29}_{-0.29}$ \\
{} & 3073-GS & $589^{+52}_{-52}$ & $0.00^{+0.00}_{-0.00}$ & $10.74^{+0.00}_{-0.00}$ & $1.11^{+0.03}_{-0.03}$ &   $9.30^{+0.00}_{-0.00}$ \\
{} & 4325-GS & $217^{+104}_{-104}$ & $1.08^{+0.07}_{-0.07}$ & $10.63^{+0.05}_{-0.05}$ & $2.04^{+0.04}_{-0.04}$ &   $12.14^{+0.07}_{-0.07}$ \\
{} & 17541-GS & $86^{+72}_{-72}$ & $0.16^{+0.06}_{-0.06}$ & $9.31^{+0.21}_{-0.43}$ & $1.32^{+0.24}_{-0.58}$ &   $10.92^{+0.28}_{-0.28}$ \\
SMC & 21698-GS & $110^{+80}_{-80}$ & $0.65^{+0.08}_{-0.08}$ & $9.81^{+0.09}_{-0.11}$ & $1.53^{+0.16}_{-0.25}$ &   $11.62^{+0.12}_{-0.12}$ \\
{} & 30423-GS & $1000^{+0}_{-0}$ & $0.06^{+0.08}_{-0.08}$ & $10.30^{+0.02}_{-0.02}$ & $-1.28^{+0.34}_{-\infty}$ &   $9.48^{+0.26}_{-0.26}$ \\
{} & 19571-GN & $515^{+268}_{-268}$ & $0.16^{+0.03}_{-0.03}$ & $10.29^{+0.04}_{-0.04}$ & $1.51^{+0.09}_{-0.11}$ &   $11.21^{+0.12}_{-0.12}$ \\
{} & 28483-GN & $1^{+0}_{-0}$ & $2.34^{+0.00}_{-0.00}$ & $10.33^{+0.00}_{-0.00}$ & $3.05^{+0.08}_{-0.10}$ &   $12.26^{+0.01}_{-0.01}$ \\
{} & 29111-GN & $147^{+46}_{-46}$ & $2.26^{+0.01}_{-0.01}$ & $11.41^{+0.02}_{-0.02}$ & $3.21^{+0.00}_{-0.00}$ &   $13.29^{+0.02}_{-0.02}$ \\
\enddata
\end{deluxetable*}

\section{An example of physical parameter estimations}\label{apd:Example_bootstrap}
\begin{figure*}[tpb]
\centering
\includegraphics[width=2.0\columnwidth, angle=0]{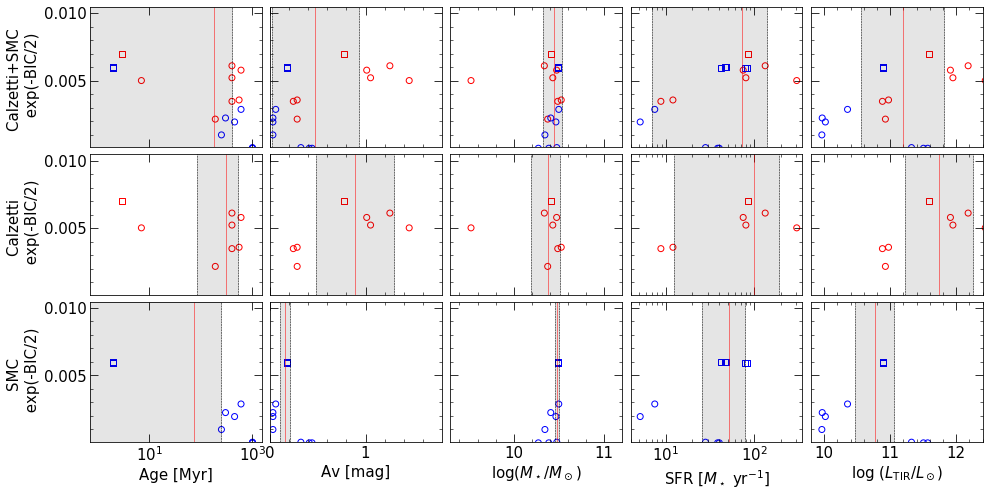}
\caption{Distribution of best-fitting physical parameters and the weights ($\exp(-{\rm BIC}/2)$) derived from SED fitting in Section \ref{subsec:SEDfit} and \ref{subsec:two-comp} for a galaxy in the final sample (1411-GS).
The red (blue) symbols represent the results by assuming Calzetti (SMC) attenuation law for the dust attenuation, and circles (squares) represent the results with one- (two-)component SED fitting.
For one-component SED fitting given in Section \ref{subsec:SEDfit}, all the results with seven SFHs are shown, while for two-component SED fitting, results that do not meet either of the criteria \ref{Criteria1} or \ref{Criteria2} in Section \ref{subsec:two-comp} are not shown.
Top row contains both results by assuming Calzetti and SMC law, and middle and bottom row show results with Calzetti and SMC law, respectively.
From left to right, physical parameters of age, $A_V$, stellar mass, SFR, and the total infrared luminosity are shown.
The vertical red solid lines and shaded regions show the best estimation (weighted mean) of each physical parameter and their uncertainties (weighted standard deviation), respectively.
}
\label{fig:bootstrap_1411}
\end{figure*}

In this appendix, we show an example of physical parameter estimation described in Section \ref{subsec:physpara} using a galaxy in the final sample (1411-GS).
Figure \ref{fig:bootstrap_1411} shows the distribution of the best-fitting physical parameter versus the weights ($\exp(-{\rm BIC}/2)$) derived by SED fittings in Section \ref{subsec:SEDfit} and \ref{subsec:two-comp} under each assumption of dust attenuation laws.
Panels in the top row contain both results of SED fitting with assuming Calzetti and SMC law.
Those in the middle (bottom) row contain only the results of SED fitting with assuming Calzetti (SMC) law.
Note that, in these panels, the results of two-component model SED fitting that do not meet the criteria \ref{Criteria1} or \ref{Criteria2} (Section \ref{subsec:two-comp}) are not shown, and are not used in the following analysis.
The red solid line in the figure shows the weighted mean value, and the shaded region presents their uncertainties obtained from the weighted standard deviations.

\section{Comparisons of results with non-parametric SFH approach}\label{apd:DB}
In this paper, we calculated the weighted means of best-fit SED models to obtain the best estimations for physical parameters, taking various SFHs into account.
In this appendix, we utilize a public SED fitting code using a non-parametric SFH approach, {\tt dense basis} \citep{iyer_2017,iyer_2019}, and examine if the main results in this paper persist when {\tt dense basis} method is adopted.
We perform SED fittings using {\tt dense basis} on galaxies in the final sample to obtain the PDFs of each physical parameter including the SFR, and followed the same procedure to calculate the SFRF, best-fitting ${\rm SFR^{\star}}$, and lower limit of SFRD from the PDFs obtained with {\tt dense basis}.
For a fair comparison, we assumed flat prior distribution in {\tt dense basis} configuration and used Calzetti attenuation law since the Calzetti law is implemented in {\tt dense basis}.

\begin{figure*}[tpb]
\centering
\includegraphics[width=2.0\columnwidth, angle=0]{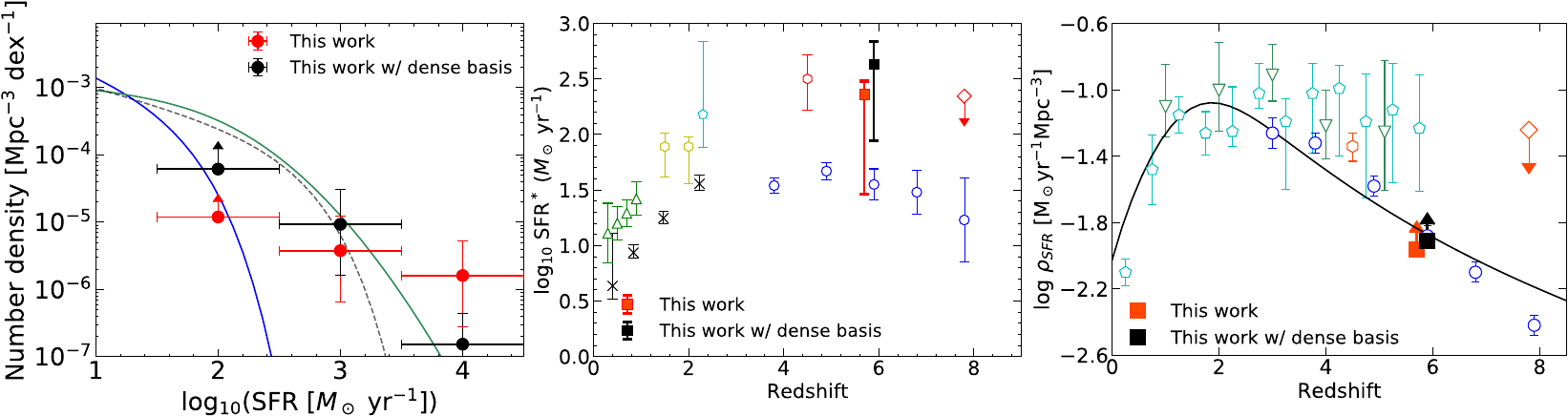}
\caption{
Comparisons of results in this paper when the weighted-mean estimation and a non-parametric SFH SED fitting code ({\tt dense basis}) is used.
Left, middle, and right panel show the resulting SFRF, ${\rm SFR}^*$, and SFRD estimation at $z\sim5.8$ in this paper, respectively.
Symbols and lines are the same as in corresponding figures (left: Figure \ref{fig:SFRF}, middle: Figure \ref{fig:SFRFparam_evo}, and right: Figure \ref{fig:SFRD}) except for filled black circles or squares, which show the result using {\tt dense basis} method.
In the middle and right panel, results in this paper at $z\sim5.8$ is displaced by $\Delta z=\pm0.1$ for clarity.
}
\label{fig:comparisons_bsdb}
\end{figure*}

Figure \ref{fig:comparisons_bsdb} shows the result of comparisons.
The resulting SFRF at $z\sim5.8$ (left panel) agrees with each other within their uncertainties, although the number density in the highest SFR bin may be overestimated with the weighted-mean estimations.
This agreement results in a persistence of estimations of the best-fitting ${\rm SFR^\star}$ value (middle panel) and lower limit on SFRD (right panel).
The best-fitting ${\rm SFR^\star}$ value is $\log_{10}({\rm SFR^\star}) = 2.36^{+0.12}_{-0.90}$ and the lower limit on the SFRD is $\log_{10}(\rho^{\rm lim}_{\rm SFR})>-1.96$ with weighted-mean method, while they are $\log_{10}({\rm SFR^\star}) = 2.63^{+0.21}_{-0.69}$ and $\log_{10}(\rho^{\rm lim}_{\rm SFR})>-1.91$ with {\tt dense basis} method.

\section{${\rm SFR^\star}$ calculation in literature}\label{apd:Schechter_fit}
The ${\rm SFR^\star}$ by \citet{bell_star_2007}, S16, and A21 in Figure \ref{fig:SFRFparam_evo} are given in the literature, but other previous studies do not give ${\rm SFR^\star}$.
Thus we derive ${\rm SFR^\star}$ by ourselves as follows.

\citet{sobral_large_2013} gave the best-fit Schechter parameters for (dust-corrected) H$\alpha$ LFs.
Since the H$\alpha$ luminosity can be converted into the SFR, we simply convert the characteristic luminosity ($L^\star_{{\rm H}\alpha}$) to the characteristic SFR (${\rm SFR^\star}$) using the following equation by assuming Chabrier03 IMF:
\begin{equation}
    L_{{\rm H}\alpha} = 2.1\times10^{41} \frac{\rm SFR}{M_\odot\ {\rm yr}^{-1}}\ {\rm erg\ s^{-1}}
\end{equation}

\citet{magnelli_evolution_2011} measured a total FIR LF, but ${\rm SFR^\star}$ for the SFRF is not derived.
We thus convert the total FIR LF into an SFRF using the following equation and obtain the best-fit Schechter parameters including ${\rm SFR^\star}$:
\begin{equation}
    L_{\rm TIR} = 1.0\times10^{10} \frac{\rm SFR}{M_\odot\ {\rm yr}^{-1}}\ L_\odot
\end{equation}

\citet{reddy_multiwavelength_2008} measured rest-frame UV, H$\alpha$, and IR LFs, but the total SFRF is not derived.
We use the total IR LF and obtain the best-fit Schechter parameters as well.

A22 derived a constraint on the H$\alpha$ LF via Schechter function.
Specifically, they obtained a forbidden region in the parameter space of the H$\alpha$ luminosity versus the number density, and derived a constraint on the Schechter parameters of the H$\alpha$ LF as follows.
They first assumed a relation between dust-uncorrected (observed) rest-frame UV luminosity ($L_{\rm UV,obs}$) and dust-corrected (intrinsic) rest-frame UV luminosity ($L_{\rm UV,int}$) as
\begin{equation}
    L_{\rm UV,int} = \eta L_{\rm UV,obs}
\end{equation}
and this correction factor $\eta$ depends on the rest-frame UV luminosity:
\begin{equation}
    \eta = a (L_{\rm UV,obs})^b
\end{equation}
With this assumption, the characteristic luminosity of the dust-corrected rest-frame UV LF ($L^\star_{\rm int}$) can be expressed using that of dust-uncorrected rest-frame UV LF ($L^\star$):
\begin{eqnarray}
    L^\star_{\rm int} &=& a (L^\star)^{b+1} \label{eqn:Lstar_int}
\end{eqnarray}
A22 derived an upper limit on the H$\alpha$ luminosity function by converting this dust-corrected rest-frame UV LF into a H$\alpha$ LF.
In this work, we use the same $L^\star$ as by A22 and their resulting parameters $(a,b)$ to calculate the upper limit on $L^\star_{\rm int}$ by Equation (\ref{eqn:Lstar_int}), and convert it into an upper limit on the ${\rm SFR^\star}$ using the following equation:
\begin{equation}
    L_{\rm UV} = 1.3 \times 10^{28} \frac{\rm SFR}{M_\odot\ {\rm yr}^{-1}}\ {\rm erg\ s^{-1}\ Hz^{-1}}
\end{equation}

\section{Conversion factor between the radio luminosity and SFR}\label{apd:SFR_rad}
In this appendix, we derive the conversion factor between the radio luminosity and the SFR (Equation \ref{eqn:radioSFR}).
\citet{condon_radio_1992} derived a conversion factor from (non-thermal) radio luminosity $L_{\rm NT}$ to the SFR for massive stars ($M_\star>5M_\odot$) assuming extended Miller-Scalo IMF.

For a fair comparison, we calculate the conversion factor to the SFR ($M_\star>0.08M_\odot$) with Chabrier03 IMF.

We start with Equation (18) by \citet{condon_radio_1992}, where the non-thermal radio luminosity $L_{\rm NT}$ and the supernova rate $\nu_{\rm SN}$ is linked:
\begin{equation}
    \left( \frac{L_{\rm NT}}{10^{22}\ {\rm W\ Hz^{-1}}} \right) \sim 13 \left( \frac{\nu_{\rm rest}}{\rm GHz}\right)^{-\alpha} \left( \frac{\nu_{\rm SN}}{\rm yr^{-1}} \right)
\end{equation}
where $\nu_{\rm rest}$ is rest-frame frequency.
We can then expect a relation between $\nu_{\rm SN}$ and SFR as
\begin{equation}
    \left(\frac{\nu_{\rm SN}}{{\rm yr}^{-1}} \right) = a \left(\frac{{\rm SFR} }{M_\odot\ {\rm yr}^{-1}} \right)
\end{equation}
where $a$ is a constant factor.
The constant factor $a$ can be calculated using an IMF $\psi(M)(\equiv dN/dM$)
\begin{dmath}
    a = \left. \left(\int_{M_{\rm SN}}^{M_U} \psi(M) dM\right) \middle/ \left(\int_{M_L}^{M_U} M\psi(M)dM \right) \right. \label{eqn:const}
\end{dmath}
where $M_L$ and $M_U$ is the lower and upper limit of the stellar mass range considered in this work ($M_L = 0.08\ M_\odot,\ M_U = 120\ M_\odot$), and $M_{\rm SN}$ is the lower bound of the stellar mass to be radio-emitting supernova remnants ($M_{\rm SN}\sim8.0\ M_\odot$), respectively.
Using $x\equiv\log M$ instead of $M$ as the variable, Equation (\ref{eqn:const}) can be written as
\begin{eqnarray}
    a = \left. \left(\int_{x_{\rm SN}}^{x_U} \xi(x) dx\right) \middle/ \left(\int_{x_L}^{x_U} \xi(x) 10^x dx \right) \right. \label{eqn:const_w_x}
\end{eqnarray}
where $\xi(x) \equiv dN/dx$.
In this work, we use Chabrier03 IMF:
\begin{dmath}\label{eqn:C03_IMF}
    \xi(\log M) = \left\{
\begin{array}{ll}
0.158\exp\left[ -\frac{(\log M - \log(0.079))^2}{2\times(0.69)^2} \right] & (M \leq 1.0\ M_\odot)\\
4.4\times10^{-2} M^{-1.3} & (M \geq 1.0\ M_\odot)
\end{array}
\right.
\end{dmath}
From Equation (\ref{eqn:const_w_x}) and (\ref{eqn:C03_IMF}), we obtain $a = 1.16\times10^{-2}\ M_\odot^{-1}$, which result in Equation (\ref{eqn:radioSFR}).


\bibliography{my_library}{}
\bibliographystyle{aasjournal}



\end{document}